\documentclass{aa}  
\usepackage{graphicx}
\usepackage{txfonts}
\usepackage{hyperref}
\hypersetup{colorlinks=true,linkcolor=blue,urlcolor=blue,citecolor=blue}
\usepackage[usenames,dvipsnames]{xcolor}
\usepackage{amsmath}
\usepackage{mathtools}
\usepackage{booktabs} 
\usepackage{soul}
\usepackage{tikz}
\usetikzlibrary{shapes.misc}
\usetikzlibrary {arrows.meta}
\usepackage{pgfplots}
\usetikzlibrary{intersections, pgfplots.fillbetween}
\usepackage{comment}
\usepackage{arydshln}

\def \cc    {\ifmmode{\,{\rm cm}^{-3}}\else{$\,{\rm cm}^{-3}$}\fi}
\def \cq    {\ifmmode{\,{\rm cm}^{-2}}\else{$\,{\rm cm}^{-2}$}\fi}
\def \mic   {\ifmmode{\,\mu{\rm m}}\else{$\mu$m}\fi}
\def \eccs  {\ifmmode{\,{\rm erg}~{\rm cm}^{-3}~{\rm s}^{-1}}\else{$\,{\rm erg}~{\rm cm}^{-3}~{\rm s}^{-1}$}\fi}
\def \ecc   {\ifmmode{\,{\rm erg}\,{\rm cm}^{-3}}\else{$\,{\rm erg}\,{\rm cm}^{-3}$}\fi}
\def \ecqs  {\ifmmode{\,{\rm erg}\,{\rm cm}^{-2}\,{\rm s}^{-1}\,{\rm 
             sr}^{-1}}\else{$\,{\rm erg}\,{\rm cm}^{-2}\,{\rm s}^{-1}\,{\rm sr}^{-1}$}\fi}
\def \ecss  {\ifmmode{\,{\rm erg}\,{\rm cm}^{-2}\,{\rm s}^{-1}}\else{$\,{\rm erg}\,{\rm cm}^{-2}\,{\rm s}^{-1}$}\fi}
\def \deg   {\ifmmode{^{\circ}}\else{$^{\circ}$}\fi} 
\def \pc    {\ifmmode{\,{\rm pc}}\else{$\,{\rm pc}$}\fi} 
\def \kms   {\ifmmode{\,{\rm km}\,{\rm s}^{-1}}\else{km~s$^{-1}$}\fi} 
\def \kmspc {\ifmmode{\,{\rm km}\,{\rm s}^{-1}\,{\rm pc}^{-1}}\else{km s$^{-1}$ pc$^{-1}$}\fi} 
\def \MJysr {\ifmmode{\,{\rm MJy\,sr}^{-1}}\else{$\,{\rm MJy\,sr}^{-1}$}\fi} 
\def \Kkms  {\ifmmode{\,{\rm K\,km\,s}^{-1}}\else{$\,{\rm K\,km\,s}^{-1}$}\fi}
\def \epso{\ifmmode{\overline{\varepsilon}_{\rm obs}}\else{$\overline{\varepsilon}_{\rm obs}$}\fi}
\def \utM{\ifmmode{u_{\theta,{\rm M}}}\else{$u_{\theta,{\rm M}}$}\fi}
\def \urM{\ifmmode{u_{r,{\rm M}}}\else{$u_{r,{\rm M}}$}\fi}
\def \twCO{\ifmmode{\rm ^{12}CO}\else{$\rm^{12}CO$}\fi} 
\def \thCO{\ifmmode{\rm ^{13}CO}\else{$\rm^{13}CO$}\fi} 
\def \CeiO{\ifmmode{\rm C^{18}O}\else{$\rm C^{18}O$}\fi} 
\def \twCN{\ifmmode{\rm ^{12}CN}\else{$\rm^{12}CN$}\fi} 
\def \thCN{\ifmmode{\rm ^{13}CN}\else{$\rm^{13}CN$}\fi} 
\def \HdCO{\ifmmode{\rm H_{2}CO}\else{$\rm H_{2}CO$}\fi} 
\def \twHdCO{\ifmmode{\rm ^{12}H_{2}CO}\else{$\rm^{12}H_{2}CO$}\fi} 
\def \thHdCO{\ifmmode{\rm ^{13}H_{2}CO}\else{$\rm^{13}H_{2}CO$}\fi} 
\def \twC{\ifmmode{\rm ^{12}C}\else{$\rm^{12}C$}\fi} 
\def \thC{\ifmmode{\rm ^{13}C}\else{$\rm^{13}C$}\fi} 
\def \Hp{\ifmmode{\rm H^+}\else{$\rm H^+$}\fi} 
\def \Cp{\ifmmode{\rm C^+}\else{$\rm C^+$}\fi} 
\def \Sp{\ifmmode{\rm S^+}\else{$\rm S^+$}\fi} 
\def \Op{\ifmmode{\rm O^+}\else{$\rm O^+$}\fi} 
\def \CFp{\ifmmode{\rm CF^+}\else{$\rm CF^+$}\fi}
\def \CHp{\ifmmode{\rm CH^+}\else{$\rm CH^+$}\fi}
\def \CHdp{\ifmmode{\rm CH_2^+}\else{$\rm CH_2^+$}\fi}
\def \CHtp{\ifmmode{\rm CH_3^+}\else{$\rm CH_3^+$}\fi} 
\def \SHp{\ifmmode{\rm SH^+}\else{$\rm SH^+$}\fi}
\def \SHdp{\ifmmode{\rm SH_2^+}\else{$\rm SH_2^+$}\fi}
\def \SHtp{\ifmmode{\rm SH_3^+}\else{$\rm SH_3^+$}\fi}
\def \twCHp{\ifmmode{\rm ^{12}CH^+}\else{$\rm^{12}CH^+$}\fi}
\def \thCHp{\ifmmode{\rm ^{13}CH^+}\else{$\rm^{13}CH^+$}\fi}
\def \CtH{\ifmmode{\rm C_2H}\else{$\rm C_2H$}\fi} 
\def \CthHt{\ifmmode{\rm C_3H_2}\else{$\rm C_3H_2$}\fi} 
\def \Htp{\ifmmode{\rm H_3^+}\else{$\rm H_3^+$}\fi} 
\def \COp{\ifmmode{\rm CO^+}\else{$\rm CO^+$}\fi} 
\def \HCOp{\ifmmode{\rm HCO^+}\else{$\rm HCO^+$}\fi} 
\def \HtOp{\ifmmode{\rm H_3O^+}\else{$\rm H_3O^+$}\fi} 
\def \HCfiN{\ifmmode{\rm HC_5N}\else{$\rm HC_5N$}\fi} 
\def \wat{\ifmmode{\rm H_2O}\else{$\rm H_2O$}\fi} 
\def \HdO{\ifmmode{\rm H_2O}\else{$\rm H_2O$}\fi} 
\def \OHp{\ifmmode{\rm OH^+}\else{$\rm OH^+$}\fi} 
\def \HdOp{\ifmmode{\rm H_2O^+}\else{$\rm H_2O^+$}\fi} 
\def \HtOp{\ifmmode{\rm H_3O^+}\else{$\rm H_3O^+$}\fi} 
\def \NHd{\ifmmode{\rm NH_2}\else{$\rm NH_2$}\fi} 
\def \NHtrois{\ifmmode{\rm NH_3}\else{$\rm NH_3$}\fi} 
\def \oxy{\ifmmode{\rm O_2}\else{$\rm O_2$}\fi} 
\def \HH{\ifmmode{\rm H_2}\else{$\rm H_2$}\fi}
\def \Jone{\ifmmode{\rm {(J=1--0)}}\else{{(J=1--0)}}\fi} 
\def \Jtwo{\ifmmode{\rm {(J=2--1)}}\else{{(J=2--1)}}\fi} 
\def \Jthr{\ifmmode{\rm {(J=3--2)}}\else{{(J=3--2)}}\fi} 
\def \Jfou{\ifmmode{\rm {(J=4--3)}}\else{{(J=4--3)}}\fi} 
\def \Jfiv{\ifmmode{\rm {J=4--3}}\else{{J=4--3}}\fi} 
\def \Ta{\ifmmode{\rm T_A}\else{$\rm T_A$}\fi} 
\def \Tas{\ifmmode{\rm T_A^*}\else{$\rm T_A^*$}\fi} 
\def \Tmb{\ifmmode{\rm T_{mb}}\else{$\rm T_{mb}$}\fi} 
\def \Tr{\ifmmode{\rm T_r}\else{$\rm T_r$}\fi} 
\def \Trs{\ifmmode{\rm T_r^*}\else{$\rm T_r^*$}\fi}
\def \NHt{\ifmmode{N_{\rm H}}\else{$N_{\rm H}$}\fi}
\def \NH{\ifmmode{N({\rm H})}\else{$N({\rm H})$}\fi}
\def \NH2{\ifmmode{N({\rm H}_2)}\else{$N({\rm H}_2)$}\fi}
\def \NCH{\ifmmode{N({\rm CH})}\else{$N({\rm CH})$}\fi}
\def \NHF{\ifmmode{N({\rm HF})}\else{$N({\rm HF})$}\fi}
\def \dens{\ifmmode{n_{\rm H}}\else{$n_{\rm H}$}\fi}
\def \densini{\ifmmode{n_{\rm H}^0}\else{$n_{\rm H}^0$}\fi}
\def \densfin{\ifmmode{n_{\rm H}^{\rm f}}\else{$n_{\rm H}^{\rm f}$}\fi}
\def \densSNR{\ifmmode{n_{\rm H}^{\rm SN}}\else{$n_{\rm H}^{\rm SN}$}\fi}
\def \nCO{\ifmmode{n({\rm CO})}\else{$n({\rm CO})$}\fi}
\def \nHF{\ifmmode{n({\rm HF})}\else{$n({\rm HF})$}\fi}
\def \nH2{\ifmmode{n({\rm H}_2)}\else{$n({\rm H}_2)$}\fi}
                 
\begin{document}

   \title{Supernova Kinetic Yield oN galactic Emission Tracers (SKYNET)}

   \subtitle{I. Contribution of radiative supernova remnants to Galactic N$^+$ emission}


   \author{G. Vigoureux\inst{\ref{lpens}, \ref{obs}}
    \and B. Godard\inst{\ref{lpens},\ref{obs}}
    \and A. Gusdorf\inst{\ref{lpens}, \ref{obs}}
    \and G. Pineau des Forêts\inst{\ref{ias}, \ref{obs}}
        }

\institute{
Laboratoire de Physique de l’École Normale Supérieure, ENS, Université PSL, CNRS, Sorbonne Université, 75005 Paris, France \label{lpens} \and
LUX, Observatoire de Paris, Université PSL, Sorbonne Université, CNRS, 75014 Paris, France \label{obs} \and
Institut d’Astrophysique Spatiale, Université Paris-Saclay, CNRS, 91405 Orsay, France \label{ias}
}

   \date{Received September 30, 20XX}


\abstract
{The radiative stage of a supernova remnant (SNR) constitutes approximately 90\% of its lifetime. Yet, only a few tens of radiative supernova remnants (r-SNRs) have been observed in the Milky Way, while catalogs of young SNRs in their adiabatic stage contain $\sim 300-400$ confirmed sources. This deficit reflects the absence of unambiguous tracers for the radiative phase, and stands as a major obstacle to understanding the impact of SNRs on the energetic balance and the chemical state of the interstellar medium.}
{We aim to identify spectral tracers of r-SNRs to enable their individual or statistical detection across the Galactic plane, and, more generally, to quantify the collective contribution of the Galactic r-SNR population to interstellar line emission.}
{We present SKYNET, a new predictive framework that couples a physically motivated model of individual r-SNRs, built upon a dedicated version of the Paris-Durham shock code, with a Galactic model that describes the spatial distribution of r-SNRs and the properties of the medium into which they expand. Together, these components allow SKYNET to predict the cumulative emission of Galactic r-SNRs in nearly one million spectral lines. As a first application, we compare the predictions of SKYNET with an \textit{Herschel} survey of the two fine-structure lines of singly ionized nitrogen.}
{The Galactic distribution predicted by SKYNET shows that random lines of sight toward the Galactic plane inevitably intercept multiple r-SNRs, whose cumulative emission may be associated with specific spectral tracers. SKYNET successfully reproduces the longitudinal profile of N$^+$ emission along the Galactic plane, which results from the projection on the sky of the spiral arms structure. The predicted line-ratio distribution matches the mean, the dispersion, and the tail of the observed distribution, showing that the model naturally explains the unexpectedly narrow range of physical conditions of the medium responsible for the N$^+$ emission. Comparison with the observed column densities  reveals that SKYNET accounts for $\sim 20$-25\% of the total Galactic N$^+$ content, demonstrating that radiative SNRs constitute a significant, and previously underappreciated, source of ionization in the Milky Way. The difference between the observed and predicted column densities could be due to the clustering of supernova remnants into superbubbles, which is currently not taken into account by the model, and to the contribution of non-SNR sources such as H\,II regions.}
{SKYNET offers a new framework for quantifying the cumulative emission of r-SNRs at Galactic scales that can be directly compared with observational surveys of atomic and molecular lines. The first application reveals that SNRs contribute significantly to the ionization of the ISM, highlighting a role that has largely been overlooked.}


   \keywords{ISM: supernova remnants -- ISM: kinematics and dynamics -- Galaxy: structure  -- ISM: atoms -- Shock waves -- ISM: lines and bands
               }

   \maketitle

\section{Introduction}

Supernovae (SNe) and supernova remnants (SNRs) form a backbone of the thermodynamical evolution of the interstellar medium (ISM) in the Milky Way. First, expanding SNRs generate and maintain a tenuous, hot ionized medium (HIM) that occupies a substantial fraction of the volume of the Galactic disk \citep{Cox1974,McKee1977,Cox2005}. In addition, by compressing and heating the gas, SNR-driven shocks induce phase transitions between the two thermally stable phases of the neutral ISM, the warm neutral medium (WNM) and the cold neutral medium (CNM) \citep[e.g.,][]{Hennebelle1999, Falle2020, Kupilas2021}. Second, SNRs are known to inject large amounts of mechanical energy into their surrounding environments. In the disk of Milky-Way-like galaxies, they are currently considered as the primary driver of interstellar turbulence \citep[e.g.,][]{MacLow2004, Kim2017, Brucy2020}. Finally, SNRs efficiently accelerate relativistic particles through diffusive shock acceleration \citep[e.g.,][]{Bell1978, Blandford1978, Blasi2013}, supplying Galactic cosmic rays. These cosmic rays provide a pervasive source of ionization and heating that contributes to the ionization state of the WNM and to the heating, ionization, and excitation of dense regions shielded from the ambient ultraviolet radiation field \citep[e.g.,][]{Wolfire2003, Dalgarno2006, Indriolo2015, Padovani2009, Padovani2018, Bialy2026}.

A striking paradox is that, although SNRs are expected to be widespread in the Milky Way, only a small fraction of them have been unambiguously detected. This reflects the fact that the Galactic census of SNRs relies on, and is intrinsically limited by, the physical tracers used for their identification. Young SNRs in their adiabatic stage are primarily identified through a combination of radio non-thermal synchrotron emission from shock-accelerated electrons\footnote{Most of remnants observed in radio in the Milky Way and in the Local Group are considered to be in the adiabatic Sedov-Taylor phase \citep{Frail1994, Albert2022}.} and X-ray continuum and line emission from the hot, shock-heated plasma in the remnant shell and interior \citep[e.g.,][]{Reynolds2008, Vink2012, Anderson2017}. Both radio- and X-ray-based techniques are affected by selection effects, including surface-brightness sensitivity, interstellar absorption, and confusion with other sources in the Galactic plane \citep[e.g.,][]{Vink2012,Green2025}. As a result, the most recent compilations list $\sim 300-400$ confirmed sources \citep{Ferrand2012,Green2025}, whereas population estimates based on the Galactic supernova rate predict between $\sim 1000$ and several thousand SNRs in their adiabatic phase \citep[e.g.,][]{Ball2023}. Ongoing wide-field radio surveys with sensitive facilities such as ASKAP, MeerKAT, LOFAR, and the VLA aim to alleviate this discrepancy \citep{Ball2023, Anderson2025, Tsalapatas2024, Dokara2023}.

The discrepancy becomes even more stringent if the long-lived radiative phase of SNR evolution is considered. As the terminal shock velocity decreases and the hot interior cools, X-ray emission fades or shifts to very soft energies that are efficiently absorbed in the Galactic plane. Radiative SNRs (r-SNRs) are therefore primarily identified through optical diagnostics of cooling shocks, such as H$\alpha$, [S\,II], [O\,II], [O\,III] emission and their ratios \citep[e.g.,][]{Fesen1985, Dopita2010, Kopsacheili2020, Vink2020}, through infrared fine-structure and near-infrared ionic and molecular lines tracing dense, shocked gas \citep[e.g.,][]{Reach2006, Andersen2011, Kokusho2026}, or through absorption studies of neutral and ionized species along specific lines of sight \citep[e.g.,][]{Jenkins1984, Welsh2001, Slavin2004, Ritchey2020}. However, because each of these tracers is affected by selection and confusion effects, a few tens of r-SNRs have been securely identified, despite the fact that the radiative phase is expected to dominate the overall Galactic population.

While individual r-SNRs remain difficult to identify, they may leave a detectable cumulative imprint on the diffuse ISM. In particular, absorption measurements of the fine-structure levels of neutral carbon have long revealed the ubiquitous presence of gas at anomalously high thermal pressures in the local ISM \citep{Jenkins2001, Jenkins2011}. Building on the scenario initially proposed by \citet{Bergin2004}, \citet{Godard2024b} showed that this high-pressure component naturally arises in radiative shocks driven by SNRs expanding in the WNM and demonstrated that such shocks are unavoidable and inevitably intersect the lines of sight probed by \citet{Jenkins2011}. Further analyses showed that the neutral-carbon line profiles can be used to constrain both the magnetic-field strength of the ambient medium and the terminal shock velocities along specific lines of sight (see Fig. 1.29 of \citealt{Godard2025}). The derived shock velocities span the range $\sim 30-150$~\kms, corresponding to SNR-evolutionary stages that are largely invisible to classical radio and X-ray diagnostics. By exploiting the statistics of many independent sightlines rather than relying on the detection of individual objects, this approach opens a new window on the late, radiative phases of SNR evolution.

Models of SNRs have a long lineage. The modern theoretical framework traces back to the similarity solutions for nuclear explosions \citep{Taylor1950,Sedov1959}. Early analytic and semi-analytic work then established the basic evolutionary sequence and the transition to the radiative stage \citep[e.g.,][]{Cox1972, McKee1977, Cioffi1988, Bandiera2004}, and provided templates for the ejecta-dominated and Sedov–Taylor phases \citep{Chevalier1982, Truelove1999}. In parallel, numerical codes evolved from one-dimensional hydrodynamics and magneto-hydrodynamics (MHD) models of SNRs expanding in idealized media \citep[e.g.,][]{Chevalier1974, Cioffi1988, Truelove1999} to three-dimensional simulations that capture asymmetries, instabilities, and environmental complexity, and are tailored to specific remnants and realistic ejecta structures \citep[e.g.,][]{Ferrand2010, DeAvillez2012, Ferrand2019, Orlando2019, Orlando2021}. Concomitantly, the microphysics has progressed from parameterized cooling functions \citep[e.g.,][]{Bertschinger1986, Thornton1998} toward explicit time-dependent non-equilibrium ionization and radiation transport coupled to the dynamics \citep[e.g.,][]{Zhang2019, Sarkar2021}.

The rapid development of numerical models and the breadth of physical processes they now include are impressive. However, despite their many advantages, state-of-the-art simulations of SNRs remain computationally expensive, which limits systematic exploration of parameter space and prevents statistical studies of remnants at different evolutionary stages distributed across the Galaxy. In this paper, we introduce SKYNET (Supernova Kinetic Yield oN galactic Emission Tracers), a framework that couples idealized models of the thermo-chemical evolution of individual r-SNRs with a statistical description of their Galactic distribution. While models of Galactic SNR populations have previously been developed to study the production and propagation of cosmic rays and high-energy photons \citep[e.g.,][]{Blasi2012, Phan2023, Batzofin2024, Stall2025}, SKYNET is designed to quantify the cumulative imprint of SNRs on line emission and to predict their statistical signatures in Galactic surveys. As a first application, we use SKYNET to estimate the spatial and statistical properties of [N\,II] fine-structure emission across the Galactic disk and compare the predictions with observations from the \textit{Herschel} Space Observatory \citep{Goldsmith2015}.

The individual model of r-SNR is presented in Sect.~\ref{sec:SNR_model}. The Galactic framework prescribing their spatial distribution and the properties of the ambient medium in which they expand is described in Sect.~\ref{sec:galac_distribution}. Model predictions for the fine-structure emission lines of N$^+$ are presented  and compared with observations in Sect.~\ref{sec:app_to_N+}. The limitations of the model and prospects for future improvements are discussed in Sect.~\ref{sec:discussion}. Section~\ref{sec:ccl} summarizes the main conclusions and the broader physical implications of this work.

\section{Model of a single radiative SNR}
\label{sec:SNR_model}

\subsection{Radiative stage of a SNR}

\begin{figure}
\centering
\includegraphics[width = \linewidth ]{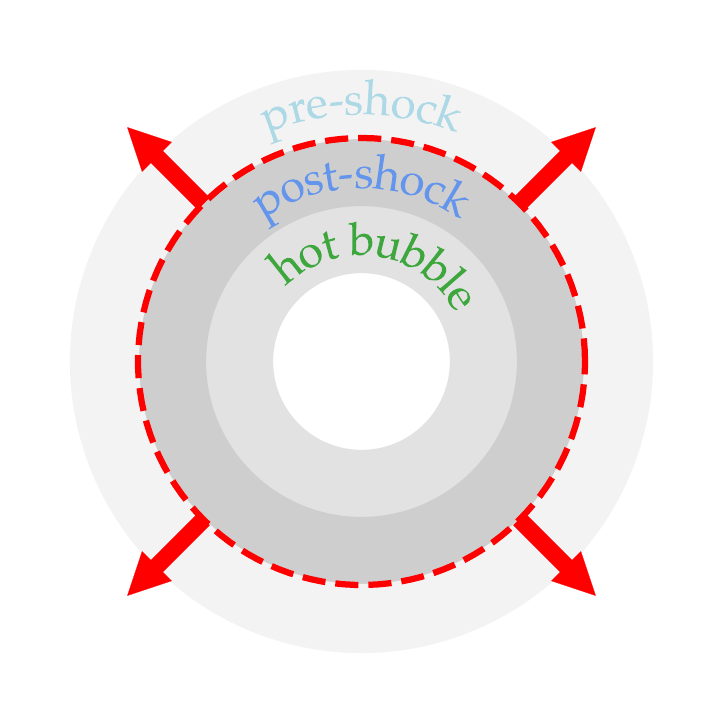}
\caption{Schematic view of the three-component model of an r-SNR. The red dashed line marks the terminal shock front and the radius of the r-SNR. The external area corresponds to the pre-shock medium, the layer immediately inside the shock front is the post-shock region, and the inner zone represents the hot bubble. The grey shading indicates the relative density of the different regions. The illustration is not to scale, as the post-shock gas and the hot bubble reside in thin shells, whereas the pre-shock medium extends over a size comparable to the radius of the r-SNR.}
\label{fig:onion}
\end{figure}

To describe the evolution of a single supernova remnant, we adopt the classical picture of a spherical blast wave expanding into a homogeneous medium. Within this idealized framework---and as detailed in Appendix~\ref{app:phase_SNR}---the temporal evolution of the terminal shock radius and velocity is governed by the explosion energy $E_{\rm SN}$, the ejecta mass $M_{\rm ej}$, the proton density of the surrounding medium $\densini$, and the time elapsed since the explosion $t$. This evolution proceeds through four distinct phases: an initial free--expansion stage ($0 \leqslant t < t_{\rm ST}$), followed by the adiabatic Sedov--Taylor phase ($t_{\rm ST} \leqslant t < t_{\rm PD}$), and ultimately the pressure--driven ($t_{\rm PD} \leqslant t < t_{\rm MC}$) and momentum--conserving phases ($t_{\rm MC} \leqslant t < t_{\rm fade}$), during which radiative losses progressively decelerate the expansion.

Throughout this work, we focus exclusively on the radiative stages of the evolution—referred to as radiative SNRs or r-SNRs—which encompass both the pressure-driven and momentum-conserving phases ($t \geqslant t_{\rm PD}$). As illustrated in Fig.~\ref{fig:onion}, an r-SNR during these stages consists of three distinct components: (i) the hot bubble, a layer of  gas at high temperature corresponding to the material shocked during the Sedov–Taylor phase and gradually cooling over time; (ii) the post-shock layer, composed of gas that is currently shocked by the blast wave and cools as it flows downstream behind the shock front; and (iii) the pre-shock gas, which corresponds to the surrounding material ionized and heated by the photons emitted from both the post-shock region and the hot bubble.

\subsection{The Paris-Durham shock code}

\begin{table}
\caption{Elements included in the model and default initial fractional elemental abundances, [X]/[H]. Numbers in parenthesis are power of 10. Bullets indicate elements that have been added relative to \citet{Godard2024a}.}
\label{tab:elem}
\begin{tabular}{l l l l l l l}
\hline
\multicolumn{2}{l}{element}   & fractional & gas       & PAHs       & grain       & ref \\
\multicolumn{2}{l}{}          & abundance  & phase     &            & cores       &     \\
\hline
H   &      & 1.00       & 1.00      & 1.8 (-7)   &             &     \\
He   &     & 1.00 (-1)  & 1.00 (-1) &            &             & a,c \\
C    &     & 3.02 (-4)  & 1.38 (-4) & 5.4 (–7)   & 1.63 (–4)   & a,c \\
N    &     & 7.94 (-5)  & 7.94 (-5) &            &             & a,c \\
O    &     & 4.42 (-4)  & 3.02 (-4) &            & 1.40 (-4)   & a,c \\
F    & $\bullet$         & 3.63 (-8)  & 3.63 (-8) &            &             & b   \\
Ne   & $\bullet$        & 8.51 (-5)  & 8.51 (-5) &            &             & b   \\
Na   & $\bullet$        & 1.74 (-6)  & 1.74 (-6) &            &             & b   \\
Mg   &     & 3.98 (-5) & 2.80 (-6) &            & 3.70 (-5)   & a,b \\
Al   & $\bullet$        & 2.82 (-6)  & 2.82 (-6) &            &             & b   \\
Si   &     & 3.67 (-5)  & 3.00 (-6) &            & 3.37 (-5)   & c   \\
P    & $\bullet$         & 2.57 (-7)  & 2.57 (-7) &            &             & b   \\
S    &     & 1.86 (-5)  & 1.86 (-5) &            &             & a,c \\
Cl   & $\bullet$        & 3.16 (-7)  & 3.16 (-7) &            &             & b   \\
Ar   & $\bullet$        & 2.51 (-6)  & 2.51 (-6) &            &             & b   \\
K    & $\bullet$         & 1.07 (-7)  & 1.07 (-7) &            &             & b   \\
Ca   & $\bullet$        & 2.19 (-6)  & 2.19 (-6) &            &             & b   \\
Fe   &     & 3.23 (-5)  & 1.50 (-8) &            & 3.23 (-5)   & a,c \\
\hline
\end{tabular}
\tablefoot{References: (a) \citep{Flower2003}, (b) \citep{Asplund2009}, (c) \citep{Kristensen2023}, leading to a PAH fractional abundance of $10^{-8}$, and a grain fractional abundance of $6.9\times 10^{-11}$ assuming a MRN distribution of grains.}
\end{table}

The out-of-equilibrium thermochemical state of each component, from the onset of the radiative phase to the late stages of the remnant’s evolution, is modeled using the Paris--Durham shock code. This code is a public numerical tool\footnote{Available on the ISM platform \url{https://ism.obspm.fr}} designed to compute the dynamical, thermal, and chemical structure of steady-state, plane-parallel interstellar shocks. In this work, we use the latest version of the model \citep{Godard2019,Godard2024a}, which allows for the treatment of shocks irradiated by an external UV field—set to the standard interstellar radiation field \citep{Mathis1983} and scaled by a factor $G_0$—as well as self-irradiated shocks propagating at velocities up to $1000$~\kms. This version notably includes the evolution of multi-ionized species, with their cooling computed using collisional rates from CHIANTI\footnote{Version 10.0.2 available at \url{https://www.chiantidatabase.org}} \citep{Dere1997, Dere2019, Del-Zanna2021}, as well as an exact treatment of the transfer of UV and X-ray photons generated by shocks and propagating into the radiative precursor. To improve the predictive power of the models, the list of elements has been expanded compared to \citet{Godard2024a}. The elements included, together with their default elemental abundances and their repartition between the gas and solid phases, are shown in Table~\ref{tab:elem}.

Two modifications were implemented to apply the code to the components described in the previous section. The one-dimensional nature of the model implies that the gas in the post-shock layer (see Fig.~\ref{fig:onion}) is confined and cannot escape in directions parallel to the shock front. Such a prescription suppresses the development of thermal and dynamical instabilities, which are known to occur in the wake of shocks \citep[e.g.][]{Falle2020, Raymond2020, Markwick2021}, and can lead to an artificially strong coupling between the photons emitted by the post-shock region and the hot bubble with the post-shock gas itself. To relax this assumption, we assume here that all photons solely interact with the pre-shock gas.

Because the typical thickness of the post-shock layer is always much smaller than the radius of the r-SNR (see Sect.~\ref{sec:shock} and Appendix~\ref{app:phase_SNR}), the plane-parallel geometry remains valid for this region. This is not true, however, for the pre-shock gas, which may extend over much larger distances (see Sect.~\ref{sec:shock}). The photon flux in the pre-shock is therefore computed by accounting not only for absorption by gas and dust (as in \citealt{Godard2024a}) but also for geometric dilution and limb brightening (see Appendix~\ref{app:dilution}).

\subsection{Evolution of the hot bubble} 
\label{sec:hotbb}

During the Sedov--Taylor phase, the temperature, density, and velocity fields in the SNR interior follow self-similar radial profiles dictated by spherical symmetry and adiabatic expansion \citep[e.g.][]{Vink2020}. The subsequent thermodynamical evolution of this hot material during the pressure-driven and momentum-conserving phases is, however, more complex. As the remnant continues to expand, the low-density gas in the innermost region---which cannot cool efficiently---exerts a pressure on the denser material located farther out in the interior. Conversely, the expansion of the remnant into the ambient medium generates a ram pressure that acts on the outermost layers. The combined effect of these opposing forces causes the material that was shocked during the Sedov--Taylor phase to assemble into a shell of gas whose dynamical evolution is set by the balance between internal pressure support and external deceleration. The thermodynamical evolution of this shell does not admit a simple analytical solution. Nevertheless, 3D numerical simulations \citep{Kim2015} and high-resolution 1D simulations \citep{Sarkar2021} suggest that the gas cools at an almost constant density before undergoing a sharp increase in density, while its thermal pressure decreases by several orders of magnitude.

To simplify the picture without altering the main physical features, we assume that during the radiative stage the interior of the SNR forms a homogeneous spherical shell (the hot bubble in Fig.~\ref{fig:onion}) that remains in pressure equilibrium with the ram, thermal, and magnetic pressures exerted by the surrounding medium. Within this framework, and under the frozen field approximation, the temperature $T$ and density $n$ of the hot bubble obey
\begin{equation} \label{eq:bb-t}
\frac{dT}{dt} = -\frac{T}{n}\mathcal{A} + T \left( 1 + \frac{2P_{\rm mag}}{P_{\rm th}} \right) \nabla u + \frac{T}{P_{\rm th}}\frac{dP}{dt},
\end{equation}
and
\begin{equation} \label{eq:bb-n}
\frac{dn}{dt} = \mathcal{A} - n \nabla u,
\end{equation}
with
\begin{equation}
\nabla u = -\frac{\gamma - 1}{\gamma P_{\rm th} + 2P_{\rm mag}}\mathcal{L} - \frac{1}{\gamma P_{\rm th} + 2P_{\rm mag}}\frac{dP}{dt}.
\end{equation}
Here, $P$, $P_{\rm th}$, and $P_{\rm mag}$ denote the total, thermal, and magnetic pressures of the hot bubble, respectively, $\nabla u$ is the divergence of the velocity field, $\gamma$ is the adiabatic index of the gas, $\mathcal{A}$ is the net particle production rate due to chemistry, and $\mathcal{L}$ is the cooling rate. The initial temperature and proton density of the shell depend on $E_{\rm SN}$ and $\densini$ and can be derived from mass and energy conservation at the onset of the radiative stage (see Appendix~\ref{app:initial_hb}). Its initial chemical state is computed by following the out-of-equilibrium evolution of a parcel of WNM gas suddenly heated to the initial hot bubble temperature (Eq.~\ref{eq:thb}) over a time interval $t_{\rm PD}$. Starting from these initial conditions, the above equations are integrated using the Paris--Durham shock code, which self-consistently incorporates chemical processes as well as radiative cooling and heating.

\begin{figure}
\centering
\includegraphics[width = \linewidth ]{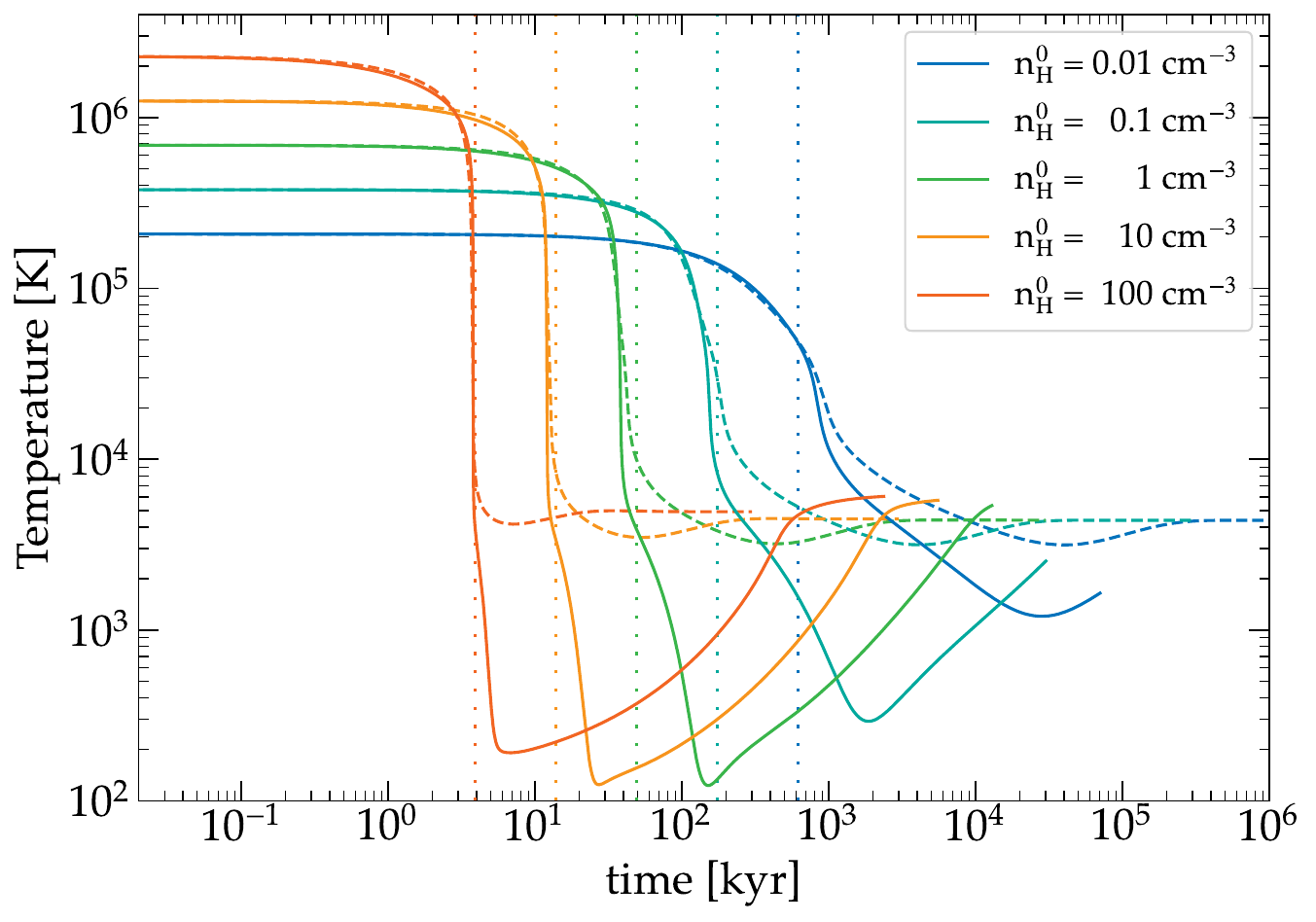}
\caption{Cooling of the hot bubble for $E_{\rm SN} = 10^{51}$\ erg and for different values of the surrounding-medium proton density $\densini$, assuming an ambient UV radiation field and a total \HH\ cosmic-ray ionization rate proportional to $\densini$,  $G_0=\densini/(1~\cc)$ and $\zeta_{\HH}=3\times10^{-16}~\mathrm{s}^{-1}\,(\densini/1~\cc)$, and a transverse magnetic field strength $B_\perp = 3~\mu\mathrm{G}\,(\densini/1~\cc)^{0.5}$ (see Sect.~\ref{sec:galac_distribution}). The temporal evolution of the temperature is shown assuming an isochoric equation of state (dashed curves) or pressure equilibrium with the surrounding medium (solid curves), as prescribed in this work. The dotted vertical lines indicate the onset time of the pressure-driven phase, $t_{\rm PD}$, derived from an analytical description of the hot bubble cooling (see Eq.~\ref{eq:tpd}).}
\label{fig:HB}
\end{figure}

The resulting thermal evolution of the hot bubble is shown in Fig.~\ref{fig:HB} for $E_{\rm SN} = 10^{51}$~erg and different values of the surrounding medium proton density \densini. Because the initial temperature increases slightly with \densini\ (see Eq.~\ref{eq:thb}), the cooling time of the hot bubble roughly scales as $\densini^{-1/2}$, in agreement with the analytical expression in Eq.~\ref{eq:tpd}. The early temperature evolution closely follows that expected for an isochoric equation of state. Once the cooling time becomes shorter than the pressure–evolution timescale, this evolution is followed by a sharp rise in density, by more than an order of magnitude, triggering a phase transition and the formation of CNM. The increase in density is ultimately stopped by the buildup of magnetic pressure. Remarkably, despite the simplicity of the underlying prescription, the resulting thermodynamical evolution is consistent with the main features obtained in high-resolution, time-dependent 1D simulations of SNRs \citep{Sarkar2021}.

\subsection{Terminal shock and pre-shock}
\label{sec:shock}

\begin{figure*}
\centering
\includegraphics[width = \linewidth ]{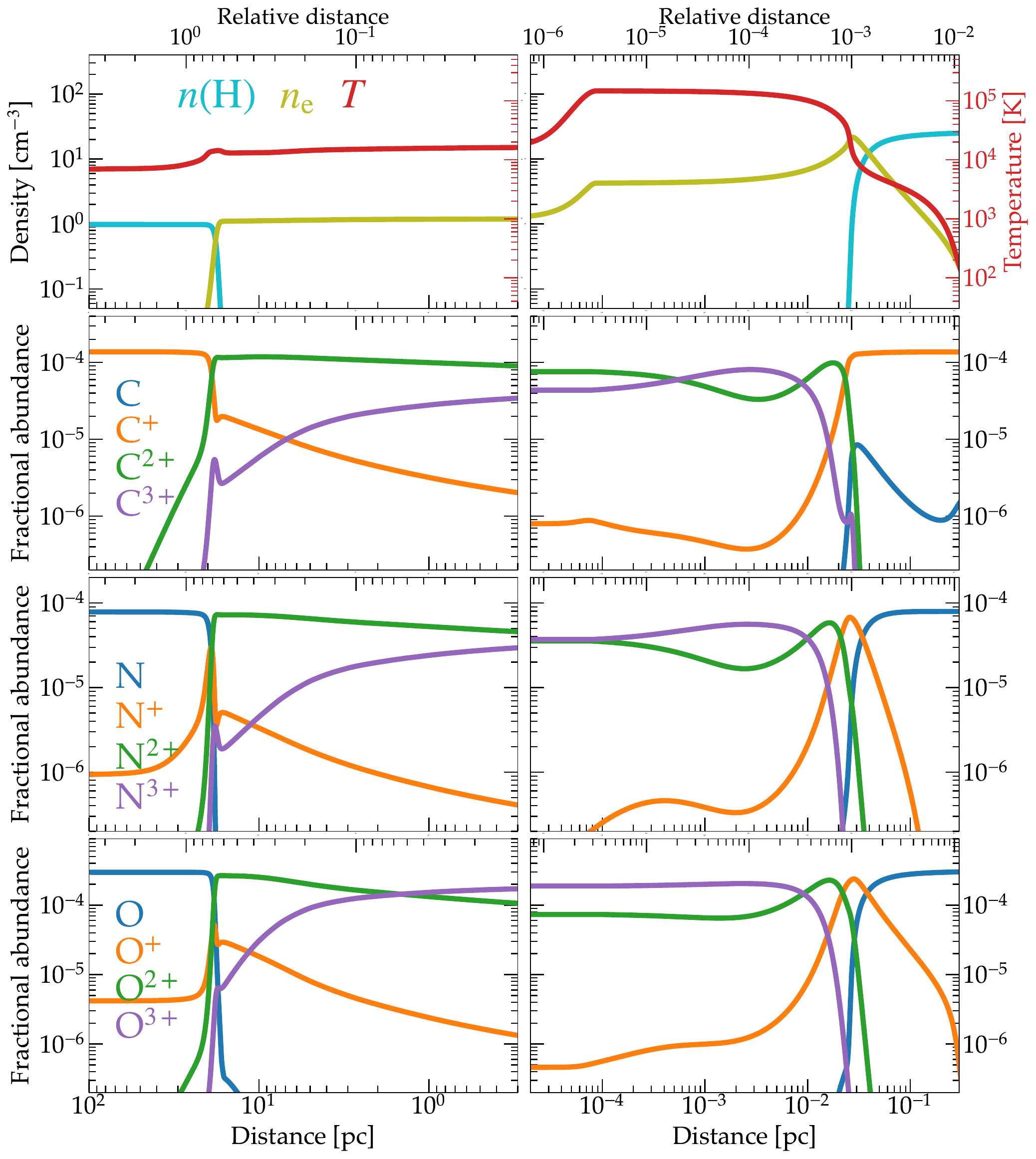}
\caption{Thermochemical structure of the pre-shock and post-shock regions for a representative r-SNR terminal shock. The profiles correspond to a shock propagating at $100~\kms$ into an ambient medium with a proton density $\densini=1~\cc$, an interstellar UV radiation field scaled by $G_0=1$, a total \HH\ cosmic-ray ionization rate $\zeta_{\HH}=3\times10^{-16}$~s$^{-1}$, and a transverse magnetic field $B_\perp=3~\mu$G. These parameters correspond to an SNR evolutionary time of $\sim 80$~kyr (see Eq.~\ref{eq:vb} and Fig.~\ref{fig:four_phases}). The temperature, neutral hydrogen density, and electron density are shown in the upper panels, while the lower panels display the relative abundances of selected carbon-, oxygen-, and nitrogen-bearing species. The left panels show the radiative precursor (pre-shock region), and the right panels show the post-shock cooling layer. In both cases, the spatial coordinate corresponds to the distance to the shock front. This distance is expressed relative to the radius of the r-SNR at the corresponding evolutionary time (Eq.~\ref{eq:rb}) on the top axis.}
\label{fig:phys_chem}
\end{figure*}

\begin{figure}[h!]
\centering
\includegraphics[width = \linewidth*29/30]{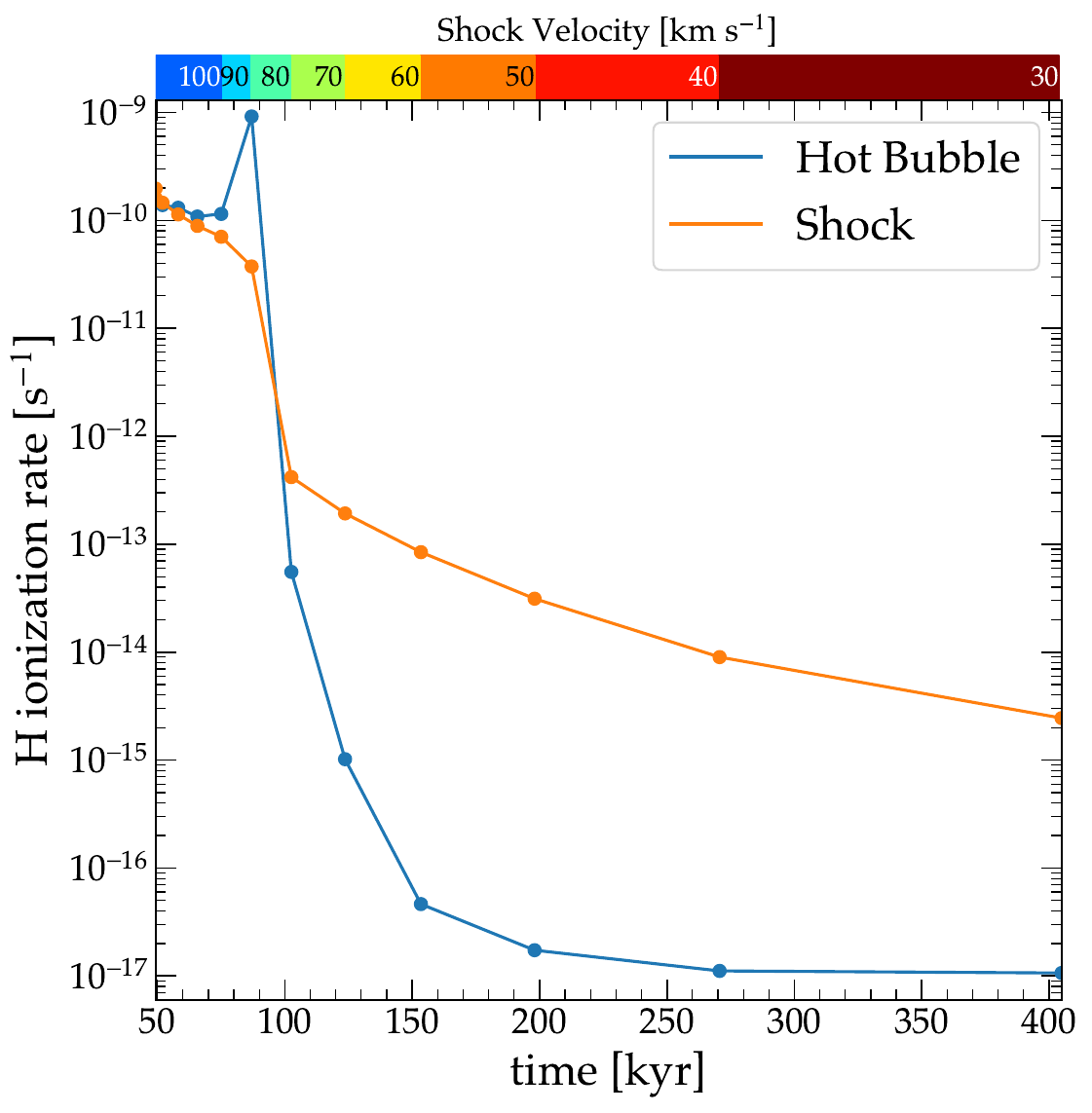}
\caption{Temporal evolution of the atomic hydrogen ionization rate at the inner border of the radiative precursor (pre-shock region) during the lifetime of a r-SNR. The model corresponds to a r-SNR with $E_{\rm SN}=10^{51}$~erg expanding into an ambient medium with a proton density $\densini=1~\cc$, an interstellar UV radiation field scaled by $G_0=1$, a total \HH\ cosmic-ray ionization rate $\zeta_{\HH}=3\times10^{-16}$~s$^{-1}$, and a transverse magnetic field $B_\perp=3~\mu$G. The photoionization rates induced by the EUV and X-ray radiation fields generated by the hot bubble and the post-shock gas are shown as blue and orange points, respectively. For reference, the corresponding temporal evolution of the terminal shock velocity, in \kms\ (Eq.~\ref{eq:vb}), is indicated on the top axis in color boxes.}
\label{fig:ionization}
\end{figure}

The physical and chemical states of the pre-shock and post-shock regions at a given evolutionary time of a r-SNR are computed following the methodology of \citet{Godard2024a}. The thermochemical evolution of a fluid particle is first followed in a Lagrangian frame during its trajectory from the pre-shock medium to the post-shock gas. The propagation of photons generated by the shock and by the hot bubble is then solved using a post-processing radiative-transfer algorithm. These two steps are repeated iteratively until convergence.

As an illustration, Fig.~\ref{fig:phys_chem} shows the converged thermochemical profiles obtained for a terminal shock propagating at $100~\kms$ into an ambient medium with a proton density of $1~\cc$ and a transverse magnetic of $3$~$\mu$G, corresponding to an SNR evolutionary time of $\sim 80$~kyr (see Eq.~\ref{eq:vb} and Fig.~\ref{fig:four_phases}). In this example, ionizing photons emitted by the post-shock gas and the hot bubble heat and ionize the pre-shock medium, leading to the formation of a radiative precursor composed of multiply ionized species. This ionized gas subsequently undergoes the classical Rankine–Hugoniot jump conditions associated with a magnetized J-type shock and eventually recombines in the cooling tail of the post-shock region. Because of the rapid, initially isobaric then isochoric, cooling of the shocked gas \citep{Godard2024a}, the post-shock region always forms a thin spherical layer at the surface of the r-SNR. In contrast, the pre-shock gas may extend over distances comparable to the r-SNR radius, as the ionizing photon flux decreases gradually through geometric dilution and absorption by the surrounding diffuse environment.

To quantify the impact of a r-SNR on the radiative precursor, Fig.~\ref{fig:ionization} displays the temporal evolution of the atomic hydrogen ionization rate induced by photons emitted by the hot bubble and the shocked gas, for a r-SNR with $E_{\rm SN}=10^{51}$~erg expanding into a medium with a proton density of $1~\cc$. Over most of the r-SNR lifetime, the resulting ionization rate at the entrance of the precursor exceeds that produced by ambient Galactic cosmic rays by at least an order of magnitude. The ionizing radiation field originates from both the hot bubble and the shocked gas, each dominating at different times. At early times, the ionization rate driven by the hot bubble and fast shocks reaches values $\gtrsim10^{-10}$~s$^{-1}$. As the hot bubble cools and the shock velocity drops below $\sim 90$~\kms, the ionization rate decreases by several orders of magnitude. This behavior is consistent with the results of \citet{Godard2024a}, who showed a sharp increase in the hydrogen ionization rate above $\sim100$~\kms\ due to the excitation of electronic levels of He$^+$ (see Fig.~12 in their paper). These results highlight the strong coupling between the dynamical evolution of the r-SNR, the cooling of the hot bubble, and the thermochemical state of the gas, and imply that the relative contribution of the different r-SNR components to ionization and emission varies significantly with time.

\section{Galactic distribution of r-SNR}
\label{sec:galac_distribution}

To assess the impact of r-SNRs on the thermochemical state of the ISM and their global emissive properties on Galactic scales, the single r-SNR model described in Sect.~\ref{sec:SNR_model} is coupled to a Galactic framework that specifies both their spatial distribution and the physical properties of the ambient medium into which they expand. The parameters controlling this Galactic model, together with their fiducial values and the range explored in this work, are summarized in Table~\ref{tab:param_galac}.

\begin{table}[h!]
\caption{Standard parameters of the full Galactic model.}
\label{tab:param_galac}
\centering
\begin{tabular}{lccl}
\hline
Parameter                              &  Symbol                 & Value     & Unit        \\
\hline
\hdashline
\multicolumn{4}{c}{\textit{Supernova remnants}} \\ \hdashline

Ejected Mass                           & $M_{\text{ej}}$         & 1.4       & $M_{\odot}$ \\
Total Energy                           & $E_{\text{SN}}$         & $10^{51}$ & erg         \\ 
Fading velocity                        & $V_{\text{fade}}$       & 30        & \kms        \\
PD onset scaling factor                & $\alpha$                & 1         &             \\
MC onset scaling factor                & $\beta$                 & 10        &             \\
\\ 
\hdashline
\multicolumn{4}{c}{\textit{Solar neighbourhood}} \\
\hdashline
Galactocentric radius                  & $R_{\odot}$             & 8.5                 & kpc         \\
Supernova scale height                 & $h_{\rm SN}^{\odot}$    & 40                  & pc          \\
Total proton density                   & $n_{\rm H}^{\odot}$     & 0.5                   & cm$^{-3}$   \\  
UV radiation field                     & $G_0^{\odot}$           & 1                   & Mathis           \\
Cosmic ray ionization rate$^{(a)}$          & $\zeta_{\HH}^{\odot}$   & $3\times 10^{-16}$  & s$^{-1}$    \\
Magnetic field parameter                & $b$       & 3                   &         \\
\\
\hdashline
\multicolumn{4}{c}{\textit{Galactic structure}} \\ 
\hdashline
Supernova rate                         & $k_{\text{SN}}$         & 46        & SN/kyr      \\
Thermonuclear fraction                 & $\delta_{\text{T}}$     & 0.3       &             \\
Minimum radius                         & $R_{\text{min}}$        & 2         & kpc         \\
Maximum radius                         & $R_{\text{max}}$        & 12        & kpc         \\
Arm number                             & $N_{\text{arm}}$        & 4         &             \\
Arm width                              & $\sigma_{\text{arm}}$   & 0.5       & kpc         \\
Arm pitch angle                        & $\theta_{\text{pitch}}$ & 13.4      & degree  \\
$\Sigma_{\rm H_2}$ flattening radius   & $R_{\text{flat}}$       & 4         & kpc         \\
$\Sigma_{\rm H_2}$ variation radius    & $R_{\text{SFR}}$        & 2         & kpc         \\
Scale height variation radius          & $R_{\text{flar}} $      & 8         & kpc         \\
HIM filling factor                     & $\phi$                  & 0.5    & kpc         \\
\hline
\end{tabular}
\tablefoot{(a) Under typical conditions within the diﬀuse neutral ISM, the total atomic and molecular cosmic ray ionization rates are related by $\zeta_{\HH} = 2.3/1.5 \zeta_{\rm H}$ \citep{Glassgold1974}.}
\end{table}

\subsection{Basic considerations}
\label{sec:basic_consid}

Following \citet{Adams2013}, the rate of supernova explosions in the Galaxy, $k_{\rm SN}$, is set to $46~{\rm SN~kyr^{-1}}$. Their population is divided into two components: a fraction $\delta_T = 30\%$ of thermonuclear supernovae (SNIa) and a fraction $(1-\delta_T)$ of core-collapse supernovae (SNII, SNIb, and SNIc). For simplicity, all supernovae are assumed to have the same ejecta mass, $M_{\rm ej} = 1.4\,M_\odot$, and the same initial kinetic energy, $E_{\rm SN}$.

Supernova explosions are distributed within a Galactic disk extending from $R_{\rm min}=2$~kpc to $R_{\rm max}=12$~kpc, with both populations sharing the same vertical distribution. Thermonuclear supernovae are distributed uniformly in azimuth, whereas core-collapse supernovae are confined to the spiral arms. Following \citet{Vallee2022}, the spiral structure is modeled as four logarithmic arms equally spaced in azimuth by $90^\circ$, with a pitch angle of $\theta_{\rm pitch} = 13.4^\circ$. The Sagittarius arm is anchored at an azimuthal angle of $50^\circ$. The distribution of core-collapse supernovae around each spiral arm in the galactic plane is finally set to follow a Gaussian profile with a dispersion $\sigma_{\rm arm} = 500$~pc \citep{Pohl1998}.

\subsection{Radial distribution of supernovae}
\label{Sect:radial}

The spatial distribution of supernovae is derived under the assumption that supernova explosions trace the star formation rate, implying that the stellar initial mass function is invariant across the Galaxy \citep{Bastian2010, Offner2014} and that stellar motions over the progenitor lifetime are negligible on Galactic scales. This latter approximation is justified for core-collapse supernovae, which originate from massive stars with  lifetimes ranging from a few to a few tens of Myr \citep{Zapartas2017}. It is, however, less appropriate for thermonuclear supernovae, whose progenitors can migrate over several hundred parsecs prior to explosion, potentially resulting in a broader vertical distribution \citep{Maoz2012, Hakobyan2017}.

Within this framework, the radial surface probability distribution function of supernova explosions in the Galactic plane, $f_{\rm SN}(R)$, and the corresponding cumulative distribution function, $F_{\rm SN}(R)$, are defined as
\begin{equation}
f_{\rm SN}(R) = \frac{\Sigma_{\rm SFR}(R)}{\int_{R_{\rm min}}^{R_{\rm max}} 2\pi r\,\Sigma_{\rm SFR}(r)\,{\rm d}r},
\end{equation}
and
\begin{equation}
F_{\rm SN}(R) = \int_{R_{\rm min}}^{R} 2\pi r\,f_{\rm SN}(r)\,{\rm d}r,
\end{equation}
where $R$ is the Galactocentric radius and $\Sigma_{\rm SFR}(R)$ is the surface density of the star formation rate at radius $R$ (in M$_{\odot}$~yr$^{-1}$~kpc$^{-2}$).

Observations of nearby and main-sequence galaxies show a strong and nearly linear correlation between the star formation rate surface density and the molecular gas surface density \citep{Leroy2013,Lin2019}, a trend that is also observed within the Milky Way \citep{Elia2022}. We therefore adopt
\begin{equation}
\Sigma_{\rm SFR}(R) \propto \Sigma_{\rm H_2}(R),
\end{equation}
so that the radial dependence of the star formation rate follows that of the molecular gas.

The radial profile of $\Sigma_{\rm H_2}(R)$ is taken from \citet{MAMD2017}, who showed that the average molecular gas surface density declines exponentially with Galactocentric radius beyond $\sim 4$~kpc. Because observational constraints are uncertain at smaller radii, we adopt a constant surface density in the inner Galaxy, chosen to ensure continuity at the transition radius. This prescription yields
\begin{equation} \label{eq:sigmaH2}
\Sigma_{\rm H_2}(R) \propto
\begin{cases}
\exp(-R_{\rm flat}/R_{\rm SFR}) & \text{for } R \le R_{\rm flat} \\
\exp(-R/R_{\rm SFR})            & \text{for } R > R_{\rm flat}
\end{cases}
    ~ , 
\end{equation}
with \(R_{\rm flat}=4\)~kpc, and \(R_{\rm SFR}=2\)~kpc\ is the characteristic variation scale of $\Sigma_{\rm H_2}$.


\subsection{Vertical distribution of supernovae}

The distributions of atomic and molecular gas indicate that the Galaxy is structured as a thin, flared disk. In particular, observations show that, at a given Galactocentric radius $R$, the vertical distribution of molecular gas is well described by a Gaussian profile with a scale height
\begin{equation} \label{eq:heightH2}
h_{\rm H_2}(R) \propto \exp\!\left(\frac{R}{R_{\rm flar}}\right),
\end{equation}
with $R_{\rm flar}=8$~kpc \citep{Nakanishi2016}. The vertical distribution of supernovae is therefore modeled as a Gaussian with a radially dependent scale height,
\begin{equation}
h_{\rm SN}(R) = h_{\rm SN}^{\odot}\,
\exp\!\left(\frac{R - R_\odot}{R_{\rm flar}}\right),
\end{equation}
where $R_\odot = 8.5$~kpc is the Galactocentric radius of the Sun. The supernova scale height at the solar radius, $h_{\rm SN}^{\odot}=40$~pc, is chosen to be consistent with the observed scale heights of very young open clusters \citep{Hao2021,Hakobyan2017} and O-B$_5$ stars \citep{Maiz-Apellaniz2001}.

\begin{figure}
\centering
\includegraphics[width = \linewidth ]{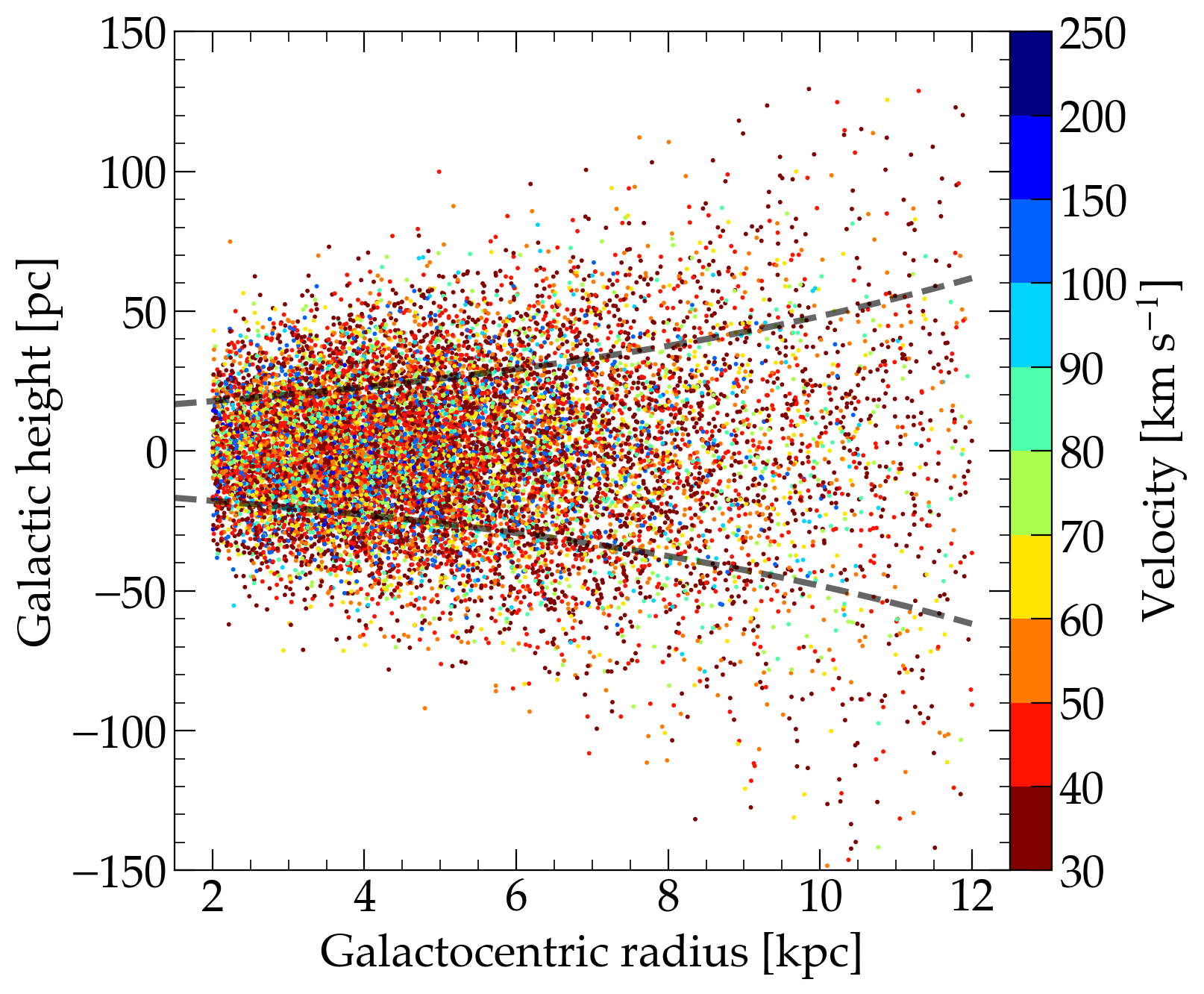}
\caption{Example realization of the spatial distribution of r-SNRs across the Galaxy. SNRs are randomly drawn from the probability distribution functions defined in the Galactic model (see main text) and are projected onto the Galactocentric radius–height plane to highlight the flaring of the disk. Each r-SNR is color-coded according to the terminal shock velocity, used as a proxy for its evolutionary stage (see Eq.~\ref{eq:vb}). 
The two gray dashed lines correspond to the standard deviation of the vertical distribution.}
\label{fig:flaring_distrib}
\end{figure}

The resulting spatial distribution of supernovae, including the flaring of the Galactic disk, is illustrated in Fig.~\ref{fig:flaring_distrib}. The enhanced concentration of remnants at small Galactocentric radii arises from the exponential radial profile of the supernova surface density, combined with the radial dependence of the vertical scale height. Most remnants are found at late evolutionary stages, which persist longer, and therefore correspond to the lowest terminal shock velocities.

\subsection{Physical conditions of the surrounding medium}
\label{sec:phys-cond}

Supernova remnants are considered to expand primarily within the diffuse phases of the interstellar medium, namely the HIM and the WNM, which together occupy most of the Galactic volume \citep[e.g.][]{Draine2011,Kim2017,Girichidis2016}. In particular, we neglect interactions between SNRs and the CNM or dense clouds, which are expected to become dynamically important when the covering fraction of dense structures over the surface of the remnant approaches unity \citep[e.g.][]{Chevalier1999,Martizzi2015,Smirnova2025}. Following \citet{Bellomi2020} (see Sect.~5.3 therein), the volume filling factor of the HIM is set to a conservative value $\phi = 0.5$. SNRs are then taken to expand freely within the HIM and to interact only with the WNM. In practice, this prescription implies that, for a given terminal shock velocity (see Eqs.~\ref{eq:rb} and \ref{eq:vb}), the effective radius of a remnant is increased by a factor (1-$\phi)^{-1/3}$ to account for the reduced volume actually filled by the WNM.

The thermochemical model of individual SNRs requires prescriptions for the physical properties of the surrounding medium, including the UV radiation field, the cosmic-ray ionization rate, the WNM proton density, and the magnetic-field strength. As discussed below, these quantities are expected, on average, to correlate with the star formation activity and thus with the distribution of molecular gas. Accordingly, each quantity is modeled with a Gaussian vertical distribution, with a scale height identical to that of the supernova population. Their values at any location in the Galaxy are therefore set by their midplane values.

Following Sect.~\ref{sec:SNR_model}, the strength of the ambient UV radiation field is parametrized by a dimensionless factor $G_0$, used as a scaling factor of the local interstellar radiation field defined by \citet{Mathis1983}. Because $G_0$ provides a proxy for the local energy density of UV photons at a given position in the Galaxy, its radial variation is expected, on average, to follow the star formation rate density. We therefore adopt the following expression for its midplane value ($z=0$)
\begin{equation} \label{eq:GC_G0}
G_0(R,0) = G_0^{\odot}
\left(\frac{\Sigma_{\rm SFR}(R)}{h_{\rm SN}(R)}\right)
\left(\frac{h_{\rm SN}(R_\odot)}{\Sigma_{\rm SFR}(R_\odot)}\right),
\end{equation}
where $G_0^{\odot}=1$ at the solar Galactocentric radius.

Cosmic rays are thought to be primarily accelerated by diffusive (Fermi) acceleration at the shock fronts of young SNRs and r-SNRs \citep[e.g.][]{Cristofari2025}. Using the prescription that the distribution of supernovae follows the star formation rate (see Sect.~\ref{Sect:radial}), the midplane value of the total cosmic-ray ionization rate of molecular hydrogen is modeled as
\begin{equation} \label{eq:GC_zeta}
\zeta_{\HH}(R,0) = \zeta_{\HH}^{\odot}
\left(\frac{\Sigma_{\rm SFR}(R)}{h_{\rm SN}(R)}\right)
\left(\frac{h_{\rm SN}(R_\odot)}{\Sigma_{\rm SFR}(R_\odot)}\right),
\end{equation}
where $\zeta_{\HH}^{\odot}=3\times10^{-16}~\mathrm{s}^{-1}$ at the solar Galactocentric radius \citep{Indriolo2015,Neufeld2017}.

\begin{figure}
\centering
\includegraphics[width = \linewidth ]{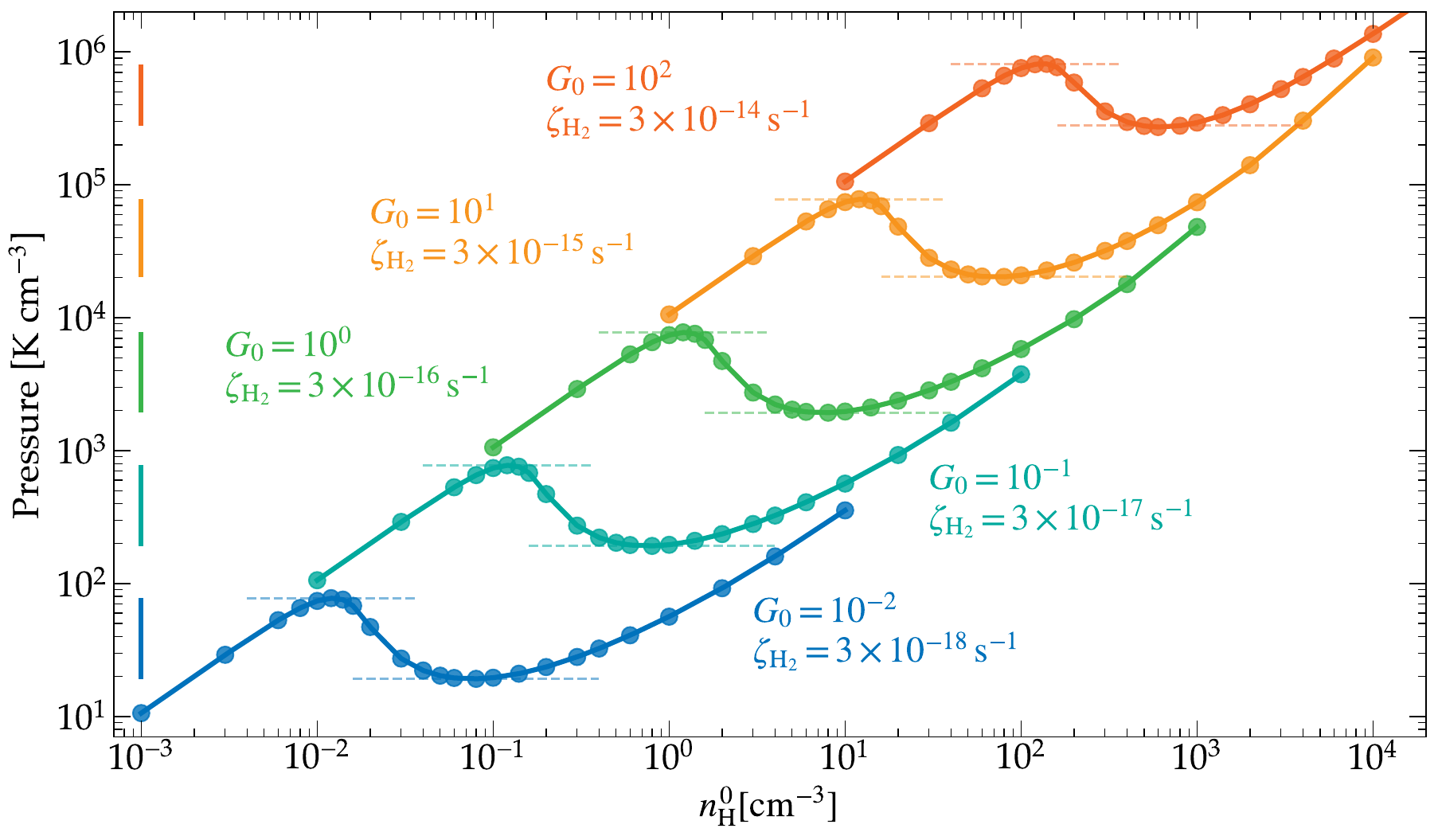}
\caption{Thermal equilibrium state of the diffuse interstellar gas obtained with the Paris-Durham code and displayed in a thermal pressure versus particle density diagram. Results are displayed for several values of $G_0$ and $\zeta_{\HH}$, assuming that both quantities are proportional, and using the gas-phase elemental abundances listed in Table~\ref{tab:elem}.For each model, the minimum and maximum thermal pressures for which a biphasic neutral medium can exist are indicated by horizontal dashed segments, while the corresponding pressure range is shown by vertical solid segments.}
\label{fig:eq_state}
\end{figure}

As illustrated in Fig.~\ref{fig:eq_state}, the fact that $G_0$ and $\zeta_{\HH}$ follow the same functional form can be exploited to model the density of the WNM. Under the constraint that $G_0$ and $\zeta_{\HH}$ are proportional to each other, thermochemical models predict that both the minimum and maximum thermal pressures of the biphasic neutral diffuse gas scale linearly with $G_0$ over the range $0.01 \leqslant G_0 \leqslant 100$. Because the WNM behaves as a thermostat at a temperature of $~8000$~K, its density is therefore expected to be proportional to $G_0$. We thus adopt:
\begin{equation} \label{eq:GC_nh0}
n_{\rm H}^0(R,0) = n_{\rm H}^{\odot}
\left(\frac{\Sigma_{\rm SFR}(R)}{h_{\rm SN}(R)}\right)
\left(\frac{h_{\rm SN}(R_\odot)}{\Sigma_{\rm SFR}(R_\odot)}\right),
\end{equation}
where $n_{\rm H}^{\odot} = 0.5$~\cc\ at the solar Galactocentric radius \citep{Wolfire2003,Draine2011}. Interestingly, this prescription implies that the WNM density follows the radial variations of the midplane density of the molecular gas (see Eqs.~\ref{eq:sigmaH2} and \ref{eq:heightH2}) as expected for a biphasic diffuse neutral medium.

Observations and modeling of the Galactic magnetic field indicate that the midplane magnetic field strength decreases with Galactocentric radius and correlates with star formation and the gas surface density \citep[e.g.][]{Beck2001,Strong2011, Beck2015,Han2018}. From a theoretical standpoint, this behavior is expected if the magnetic field is maintained in approximate equipartition with interstellar turbulence. Recent observational analyses support this picture and show that the magnetic energy density scales approximately linearly with the gas density over a wide range of conditions \citep[e.g.][]{Seta2025}. We therefore adopt a magnetic field strength perpendicular to the SNR expansion of the form
\begin{equation} \label{eq:GC_Bperp}
B_\perp(R,z) = 1\,\mu{\rm G}\,\,b \left( \frac{n_{\rm H}^0(R,z)}{1\,{\rm cm}^{-3}} \right)^{1/2}
\end{equation}
at each position $(R,z)$ in the Galaxy, with a magnetic field parameter of $b = 3$, which lies at the upper end of the relation reported by \citet{Seta2025}.

\begin{figure}
\centering
\includegraphics[width = \linewidth ]{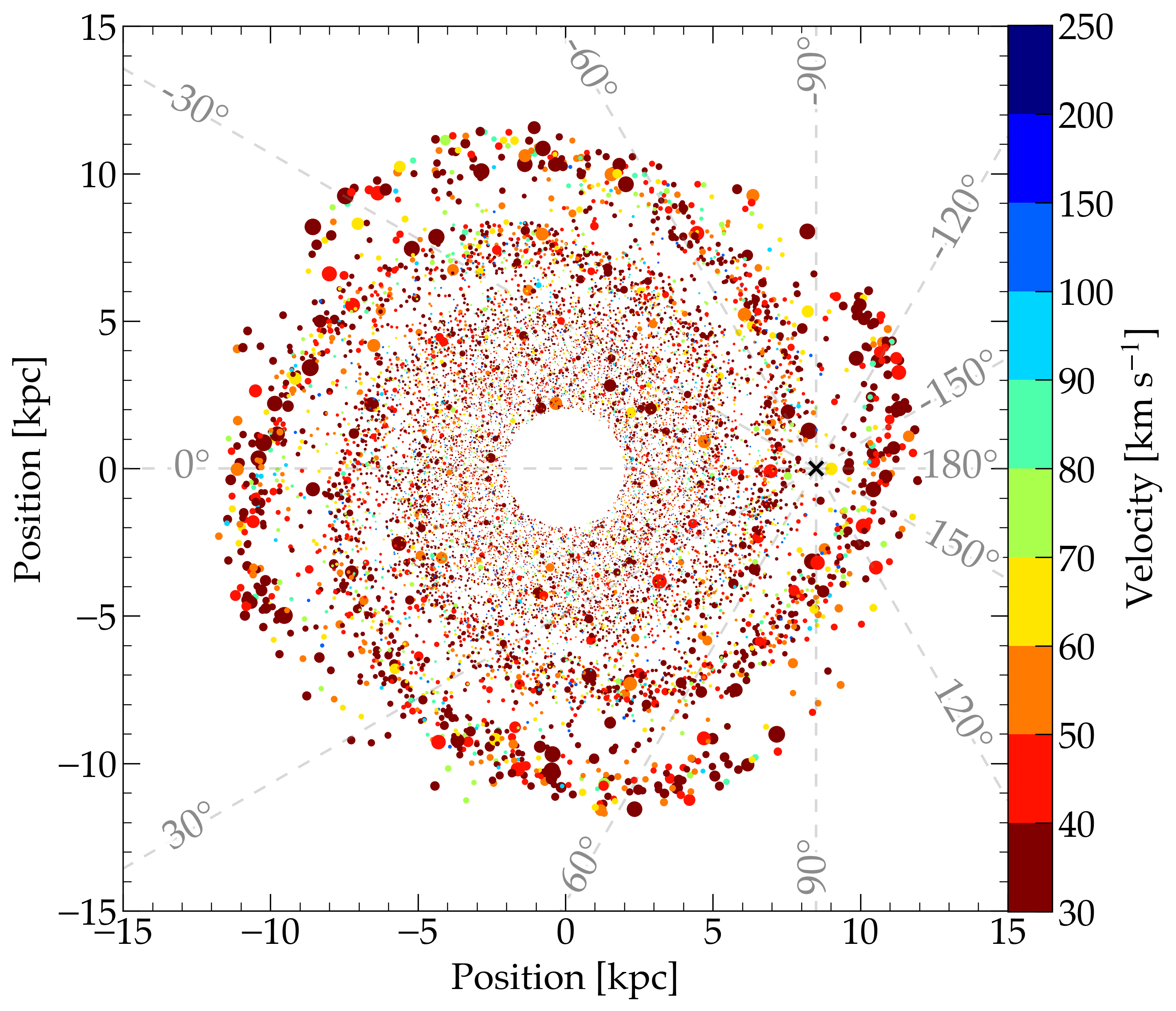}
\caption{Example realization of the face-on spatial distribution of r-SNRs across the Galaxy. This figure shows the same r-SNR population as in Fig.~\ref{fig:flaring_distrib}, projected onto the Galactocentric radius-azimuth plane. Each r-SNR is color-coded by its terminal shock velocity (see Eq.~\ref{eq:vb}). 
The figure is to scale, and the size of each symbol corresponds to the physical radius of the remnant derived from Eq.~\ref{eq:rb}. The increase in the typical r-SNR radius with Galactocentric distance reflects the adopted radial profile of the ambient gas density.}
\label{fig:SNR_galax_distrib}
\end{figure}

\begin{figure*}[h!]
\centering
\includegraphics[width = \linewidth ]{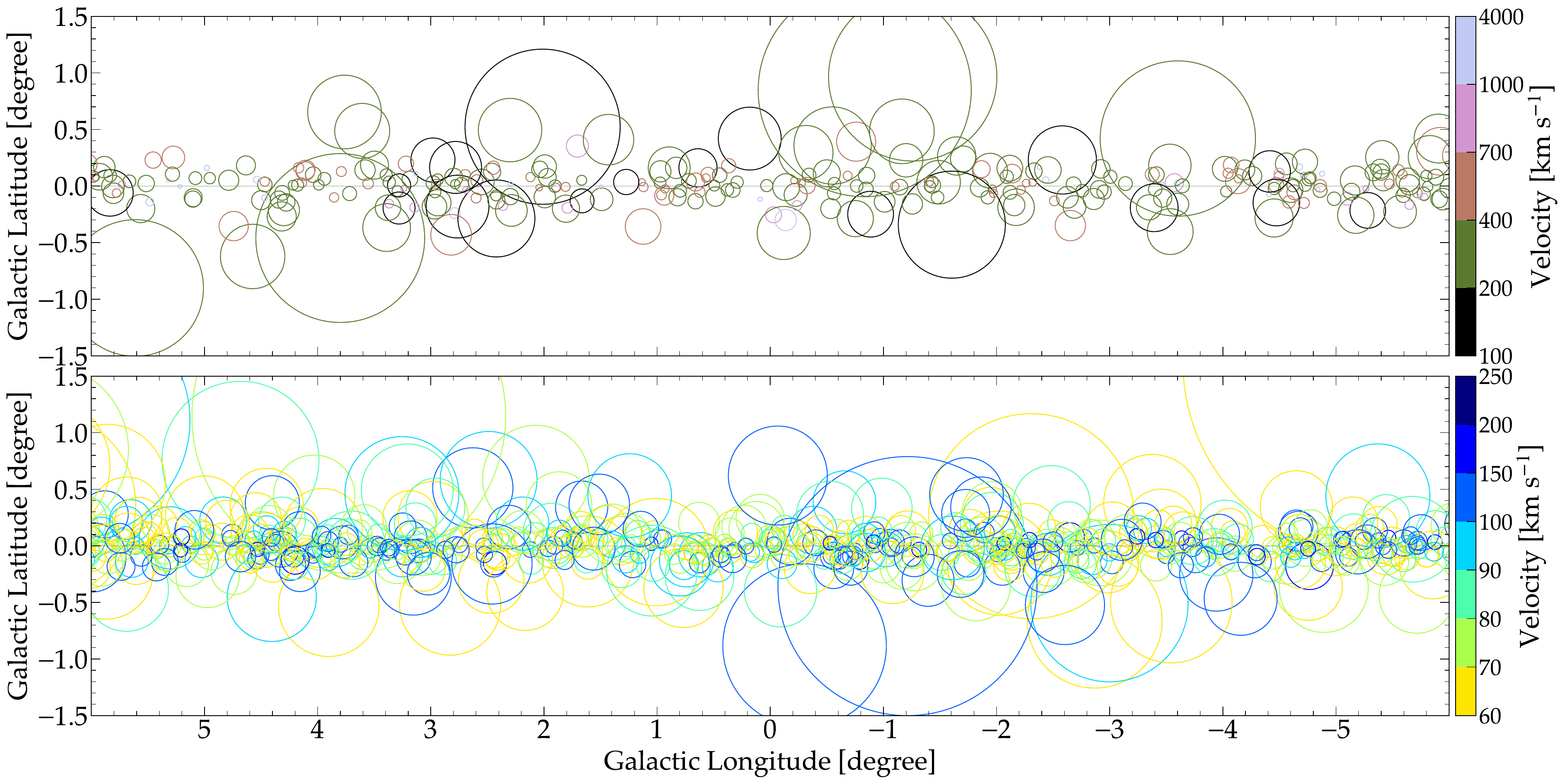}
\caption{Example realization of the angular distribution of SNRs as seen from the Solar position. Non-radiative SNRs are shown in the top panel, while radiative SNRs are shown in the bottom panel. The figure displays the same SNR population as in Fig.~\ref{fig:flaring_distrib}, projected onto the sky. Each SNR is color-coded by its terminal shock velocity (see Eq.~\ref{eq:vb}). For clarity, only SNRs with terminal shock velocities above $60~\mathrm{km,s^{-1}}$ are displayed. The figure is to scale, and the angular size of each SNR is derived from the physical radius of the remnant obtained with Eq.~\ref{eq:rb}.}
\label{fig:slice_SNR_galax_distrib}
\end{figure*}

\subsection{Total number of supernovae}
\label{subsec:snr_number}
Given the three-dimensional distribution of SNRs and the properties of the surrounding medium, the total numbers of non-radiative and radiative remnants present in the Galaxy, $\mathcal{N}_{\rm nr\mbox{-}SNR}$ and $\mathcal{N}_{\rm r\mbox{-}SNR}$, are given by
\begin{equation}
        \mathcal{N}_{\rm nr-SNR}=   k_{\mathrm{SN}} \int_{R_{\mathrm{min}}}^{R_{\mathrm{max}}}   \frac{ 2 \pi  rf_{\mathrm{SN}}(R) }{\sqrt{2\pi} h_{\text{SN}}(R)} \int_{-\infty}^{\infty}e^{-z^2/2h_{\text{SN}}(R)^2} t_{\mathrm{PD}}(R,z)\, dz\, dR
\end{equation}
and
\begin{equation}
        \mathcal{N}_{\rm r-SNR}=   k_{\mathrm{SN}} \int_{R_{\mathrm{min}}}^{R_{\mathrm{max}}}   \frac{ 2 \pi  rf_{\mathrm{SN}}(R) }{\sqrt{2\pi} h_{\text{SN}}(R)} \int_{-\infty}^{\infty}e^{-z^2/2h_{\text{SN}}(R)^2} \Delta t(R,z)\, dz\, dR,
\end{equation}
where $\Delta t(R,z)=t_{\mathrm{fade}}(R,z)-t_{\mathrm{PD}}(R,z)$ is the duration of the radiative phase (see Appendix~\ref{app:phase_SNR}). The time $t_{\mathrm{PD}}$ depends on the supernova energy and the ambient density, while $t_{\mathrm{fade}}$ also depends on the velocity dispersion of the WNM. Since the explosion energy and the WNM velocity dispersion are treated as constants, the spatial dependence of $\Delta t$ arises solely from variations in the ambient density. The standard parameters adopted in Table~\ref{tab:param_galac} yield a total number of non-radiative SNRs, $\mathcal{N}_{\rm nr-SNR} \sim 2200$, in agreement with the estimate of \citet{Ball2023}, and a corresponding number of radiative SNRs, $\mathcal{N}_{\rm r-SNR} \sim 14\,000$.


 \subsection{Face-on and angular distributions}
\label{sec:latitude_distrib}

The Galactic model described above allows the generation of random realizations of the r-SNR population across the Galaxy. An example of such a realization is illustrated in Figs.~\ref{fig:SNR_galax_distrib} and \ref{fig:slice_SNR_galax_distrib}, which display, respectively, a face-on view of the resulting Galactic r-SNR distribution and the corresponding angular distribution of both non-radiative and radiative SNRs as seen from the Solar position. These figures highlight the spiral structure of the Milky Way, the radial and vertical profile of the ambient density, and the strong concentration of remnants toward the Galactic plane.

Apart from the presence of a central bar, the resulting spiral pattern closely resembles those obtained by \citet{Ahlers2009} and \citet{Phan2023}, who modeled Galactic supernova distributions to study the generation and propagation of cosmic rays. Given the small scale height of the Galactic disk, nearly all r-SNRs are confined within $\sim 1^\circ$ of the Galactic plane, with only a few exceptions corresponding to nearby remnants. The angular distribution of non-radiative SNRs, which are expected to emit strong synchrotron radiation, is qualitatively consistent with recent observations of Galactic radio sources obtained with MeerKAT at 1.3~GHz \citep{Goedhart2024} and with the Murchison Widefield Array at lower frequencies \citep{Mantovanini2025}.


\subsection{Porosity}

Beyond their projected distributions, these realizations can be used to assess the volumetric impact of r-SNRs and the global coherence of the model. To this end, we compute the porosity factor, $q$, defined as the ratio between the total volume occupied by r-SNR interiors—regardless of their mutual overlap—and the volume of the Galaxy \citep[e.g.,][]{Slavin1993}. The Galactic volume is approximated as that of a flared disk extending out to $R_{\rm max}$, with a radius-dependent vertical thickness $2h_{\rm SN}(R)$. Taking into account only the r-SNRs that lie within this vertical extent (see Fig.~\ref{fig:flaring_distrib}) and using the parameters listed in Table~\ref{tab:param_galac} yields a porosity factor $q \simeq 0.13$.

Given the radial structure of r-SNRs, most of their volume is occupied by the hot, tenuous interior, surrounded by a comparatively thin shell of cooling material. The porosity factor is therefore directly related to the volume filling factor of the HIM, $\phi$ (see Sect.~\ref{sec:phys-cond}). However, the two quantities are not strictly equivalent. The porosity factor measures the volume occupied by remnants that are still identifiable as r-SNRs, whereas the hot cavities they create may persist after the remnants have faded, until they are replenished by the surrounding warm medium. A simple estimate therefore gives
\begin{equation}
\phi \simeq q \frac{t_{\rm ref}}{t_{\rm fade}},
\end{equation}
where $t_{\rm ref}$ is the characteristic refilling time of a faded cavity. Estimating this timescale from the remnant radius at $t_{\rm fade}$ and the one-dimensional velocity dispersion of the WNM yields $t_{\rm ref}/t_{\rm fade}\simeq 3-5$. Combined with the porosity factor derived above, this implies $\phi \simeq 0.39-0.65$, in  agreement with the fiducial value $\phi=0.5$ adopted in the model.

Early analytical models of the multiphase ISM predicted large porosity factors, with values $q \geqslant 1$ \citep[e.g.,][]{McKee1977}, reflecting the assumption of long-lived, weakly cooling hot phases. Subsequent semi-analytical studies incorporating radiative SNR evolution and more realistic ambient densities revised these estimates downward, yielding porosity factors $q \sim 0.1-0.3$ in the Solar neighborhood \citep[e.g.][]{Slavin1993}. More recent numerical simulations of SN-driven, stratified galactic disks do not quote porosity explicitly, but instead report hot gas volume filling factors in the range $\phi \sim 0.2-0.8$, depending on the supernova rate and background density \citep[e.g.][]{Walch2015,Hill2018}. The porosity value obtained here leads to a volume filling factor that falls within this modern range and below the threshold for thermal runaway identified in recent simulations \citep{Li2015}.



\section{Application to Galactic [N\,II] emission}
\label{sec:app_to_N+}

The combination of the thermochemical model of individual r-SNRs (Sect.~\ref{sec:SNR_model}) with the Galactic model (Sect.~\ref{sec:galac_distribution}) gives access to a myriad of observational tracers, including line and continuum emission, their variance across the Galaxy, and the distribution of individual line profiles along with their systemic velocities. In particular, Fig.~\ref{fig:slice_SNR_galax_distrib} shows that r-SNRs are unavoidable in the Galactic plane, and that any random line of sight at low Galactic latitude necessarily intercepts one or several structures. As a first application, we explore here the impact of r-SNRs on the production of singly ionized nitrogen and compare the predictions of the model with available observational constraints on the Galactic plane.


\subsection{Observational sample}

The observational constraints are taken from \citet{Goldsmith2015}, who performed a Galactic plane survey of the two fine-structure lines of N$^+$ at 122 and 205~$\mu$m in emission using the PACS and HIFI instruments aboard the \textit{Herschel} Space Observatory. This survey includes 147 lines of sight at zero Galactic latitude, with a longitude spacing of $\sim1^\circ$ in the inner Galaxy ($|l|\lesssim60^\circ$) and $\sim5^\circ$ at larger longitudes.

As shown by \citet{Goldsmith2015}, this dataset reveals several key observational features. Detections of at least one [N\,II] line are reported toward 118 lines of sight and show a strong dependence on Galactic longitude. Only eight detections are reported at $|l| \gtrsim 60^\circ$, while most detections arise within $-60^\circ < l < 60^\circ$. Within this range, the emission sharply increases toward the inner Galaxy and reaches a plateau characterized by large spatial variations, spanning nearly two orders of magnitude in intensity. HIFI data collected along ten lines of sight indicate that the emission is distributed in several velocity components, with systemic velocities broadly consistent with Galactic rotation. The ratio of the two fine-structure transitions provides a constraint on the electron density in the emitting regions, yielding $n_e \simeq 10$--$100\,\mathrm{cm^{-3}}$, with a mean value of $\sim 30$~\cc\ and a dispersion smaller than a factor of two. Finally, the inferred N$^+$ column densities are systematically below $10^{17}\,\mathrm{cm^{-2}}$, justifying the use of the optically thin approximation to derive column densities from the observed line intensities.

Together, these morphological, statistical, and physical properties provide strong constraints on the conditions and the origin of the [N\,II]-emitting gas. The velocity structure observed with HIFI and the large dispersion in total intensities suggest that the [N\,II] emission arises from the collection of several regions with a sky covering factor larger than unity. The dependence on Galactic longitude underlines the importance of Galactic structure in shaping the observed emission, in particular the contribution of spiral arms such as the Sagittarius arm, which is tangential to the line of sight near longitudes of $\sim \pm60^\circ$ \citep[e.g.,][]{Kachelriess2025}. Last, the narrow range of electron densities derived from the full dataset suggests that the regions responsible for [N\,II] emission share remarkably similar physical conditions throughout the Galaxy. This implies that such regions are either common, or relatively rare yet particularly efficient at producing and exciting N$^+$.


\subsection{Production and excitation of N$^+$}
\label{sec:prodN+_SNR}

\begin{figure}
\centering
\begin{tikzpicture}
    \draw [black,thick, -Stealth](1,0) -- (1,10) ;
    \draw ( -0.1,10)  node {\large \bf{$E ~ [K]$}} ;
    \draw [white,very thick](1,2.5+1.25) -- (1,2.6+1.25) ; 
    \draw [black,thick](0.75,2.55 +1.25) -- (1.25,2.45+1.25) ;
    \draw [black,thick](0.75,2.65+1.25) -- (1.25,2.55+1.25) ;
    \draw [lightgray,thin, dashed](1.5,2.5+1.25) -- (6.5,2.5+1.25) ; 

    \draw [black,thick](0.75,0.5) -- (1,0.5) ; 
    \draw ( 0.5,0.5)  node {\large \bf{$0$}} ;
    \draw [black,thick](0.9,0.7*1.5 + 0.5) -- (1,0.7*1.5 + 0.5) ; 
    \draw ( 0.7,0.7*1.5 + 0.5)  node {\tiny \bf{$70$}} ;
    
    \draw [black,thick](0.75,1*1.5 + 0.5) -- (1,1*1.5 + 0.5) ; 
    \draw ( 0.4,1*1.5 + 0.5)  node {\large \bf{$10^2$}} ;

    \draw [black,thick](0.9,1.8*1.5 + 0.5) -- (1,1.8*1.5 + 0.5) ; 
    \draw ( 0.7,1.8*1.5 + 0.5)  node {\tiny \bf{$188$}} ;

    \draw [black,thick](0.75,2*1.5 + 0.5) -- (1,2*1.5 + 0.5) ; 
    \draw ( 0,2*1.5 + 0.5)  node {\large \bf{$2\times 10^2$}} ;

    \draw [black,thick](0.75,-2/28 + 4.5) -- (1,-2/28 + 4.5) ; 
    \draw ( 0,-2/28+ 4.5)  node {\large \bf{$2\times 10^4$}} ;
    
    
    \draw [black,thick](0.75,28/28 + 4.5) -- (1,28/28 + 4.5) ; 
    \draw ( 0,28/28+ 4.5)  node {\large \bf{$5\times 10^4$}} ;


    \draw [black,thick](0.75,78/28 + 4.5) -- (1,78/28 + 4.5) ; 
    \draw ( 0.3,78/28+ 4.5)  node {\large \bf{$ 10^5$}} ;
    

    \draw [black,thick](0.75,128/28 + 4.5) -- (1,128/28 + 4.5) ; 
    \draw ( -0.1,128/28+ 4.5)  node {\large \bf{$1.5 \times 10^5$}} ;

    \draw [black,thick, -Stealth](0,0) -- (7,0) ;
    \draw ( 7,-0.5)  node {\large \bf{$l$}} ;

    \draw [black,thick](2.25,-0.25) -- (2.25,0) ;
    \draw ( 2.25,-0.5)  node {\large \bf{$0$}} ;
    \draw ( 2.25,-1)  node {\large (S)} ;
    \draw [black,thick](3.75,-0.25) -- (3.75,0) ;
    \draw ( 3.75,-0.5)  node {\large \bf{$1$}} ;
     \draw ( 3.75,-1)  node {\large (P)} ;
    \draw [black,thick](5.25,-0.25) -- (5.25,0) ;
    \draw ( 5.25,-0.5)  node {\large \bf{$2$}} ;
    \draw ( 5.25,-1)  node {\large (D)} ;

    \definecolor{mycolor111}{rgb}{0.5,0,0}
    \draw [mycolor111,thick, -Stealth,opacity = 0.85](3.75 ,135/28 + 4.5 - 0.02) -- (3.75 ,0.7*1.5 + 0.5+0.05) ; 

    \definecolor{mycolor101}{rgb}{0.62,0,0}
    \draw [mycolor101,thick, -Stealth,opacity = 0.85](3.75 ,135/28 + 4.5 - 0.02) -- (3.75 ,0.7*1.5 + 0.5+0.05) ; 
    \definecolor{mycolor102}{rgb}{0.5,0,0}
    \draw [mycolor102,thick, -Stealth,opacity = 0.85](3.75 ,135/28 + 4.5 - 0.02) -- (3.75,1.8*1.5 + 0.5+0.05) ; 

    \definecolor{mycolor90}{rgb}{0.59,0,0}
    \draw [mycolor90,thick, -Stealth,opacity = 0.85](3.75 ,135/28 + 4.5 - 0.02) -- (3.75,0.5+0.05) ; 
    \definecolor{mycolor91}{rgb}{0.62,0,0}
    \draw [mycolor91,thick, -Stealth,opacity = 0.85](3.75 ,135/28 + 4.5 - 0.02) -- (3.75 ,0.7*1.5 + 0.5+0.05) ; 
    \definecolor{mycolor92}{rgb}{0.57,0,0}
    \draw [mycolor92,thick, -Stealth,opacity = 0.85](3.75,135/28 + 4.5 - 0.02) -- (3.75,1.8*1.5 + 0.5+0.05) ; 

    \definecolor{mycolor80}{rgb}{0.68,0,0}
    \draw [mycolor80,thick , -Stealth,opacity = 0.85](5.25,110/28 + 4.5 - 0.02) -- (3.75+0.1,0.5+0.05) ; 
    \definecolor{mycolor81}{rgb}{0.71,0,0}
    \draw [mycolor81,thick, -Stealth,opacity = 0.85](5.25,110/28 + 4.5 - 0.02) -- (3.75 +0.1,0.7*1.5 + 0.5+0.05) ; 
    \definecolor{mycolor82}{rgb}{1,0.12,0}
    \draw [mycolor82,thick, -Stealth,opacity = 0.85](5.25,110/28 + 4.5 - 0.02) -- (3.75+0.1,1.8*1.5 + 0.5+0.05) ; 
    
    \definecolor{mycolor71}{rgb}{0.64,0,0}
    \draw [mycolor71,thick, -Stealth,opacity = 0.85](5.25 +0.05/2,110/28 + 4.5 - 0.02) -- (3.75 +0.1 ,0.7*1.5 + 0.5+0.05) ; 
    \definecolor{mycolor72}{rgb}{0.79,0,0}
    \draw [mycolor72,thick, -Stealth,opacity = 0.85](5.25 +0.05/2,110/28 + 4.5 - 0.02) -- (3.75 +0.1,1.8*1.5 + 0.5+0.05) ; 

    \definecolor{mycolor62}{rgb}{0.61,0,0}
    \draw [mycolor62,thick, -Stealth,opacity = 0.85](5.25,110/28 + 4.5 - 0.02) -- (3.75+0.1,1.8*1.5 + 0.5 +0.05) ; 

    \definecolor{mycolor51}{rgb}{0.62,1,0.35}
    \draw [mycolor51,thick, -Stealth,opacity = 0.85](2.25 ,45/28 + 4.5 - 0.02) -- (3.75 - 0.1,0.7*1.5 + 0.5+0.05) ; 
    \definecolor{mycolor52}{rgb}{0.7,1.0,0.26}
    \draw [mycolor52,thick, -Stealth,opacity = 0.85](2.25 ,45/28 + 4.5 - 0.02) -- (3.75 - 0.1,1.8*1.5 + 0.5+0.05) ; 

    \definecolor{mycolor41}{rgb}{0,0.88,0.98}
    \draw [mycolor41,thick, -Stealth,opacity = 0.85](2.25,25/28 + 4.5 - 0.02) -- (3.75 - 0.2 ,0.7*1.5 + 0.5+0.05) ; 
    \definecolor{mycolor42}{rgb}{0,0.3,1}
    \draw [mycolor42,thick, -Stealth,opacity = 0.85](2.25,25/28 + 4.5 - 0.02) -- (3.75 - 0.2,1.8*1.5 + 0.5+0.05) ; 
    \definecolor{mycolor43}{rgb}{0.29,1,0.68}
    \draw [mycolor43,thick, -Stealth,dashed,opacity = 0.85](2.25,25/28 + 4.5 - 0.02) -- (5.25,4.5+0.05) ; 

    \definecolor{mycolor30}{rgb}{0,0,0.7}
    \draw [mycolor30,thick, -Stealth,opacity = 0.85](5.25,4.5 - 0.02) -- (3.75 +0.2,0.5+0.05) ; 
    \definecolor{mycolor31}{rgb}{0,0.49,1}
    \draw [mycolor31,thick, -Stealth,opacity = 0.85](5.25,4.5 - 0.02) -- (3.75  +0.2,0.7*1.5 + 0.5+0.05) ; 
    \definecolor{mycolor32}{rgb}{0,0.61,1}
    \draw [color = mycolor32,thick, -Stealth,opacity = 0.85](5.25,4.5 - 0.02) -- (3.75 +0.2,1.8*1.5 + 0.5+0.05) ; 

    \definecolor{mycolor21}{rgb}{0,0,1}
    \draw [mycolor21,thick, -Stealth](3.75,1.8*1.5 + 0.5 - 0.02) -- (3.75 ,0.7*1.5 + 0.5+0.05) ; 

    \definecolor{mycolor10}{rgb}{0,0,0.87}
    \draw [mycolor10,thick, -Stealth](3.75 ,0.7*1.5  + 0.5 - 0.02) -- (3.75,0.5+0.05) ; 

    \draw ( 6.65 ,9.5)  node {\large {$\log(A_{ij} ~ [\text{s}^{-1}])$}} ;
    \begin{axis}[hide axis,scale only axis,colormap/jet,colorbar horizontal,point meta min=-7,
    point meta max=9, colorbar style={xtick={-7,-5,-3,-1,1,3,5,7,9}, xticklabel style={xshift=10pt,yshift=5pt},rotate = 90, at={(0.75,1.25)}}]
    \end{axis}
    
    \draw [black,thick](3.5,0.5) -- (4,0.5) ; 

    \draw [black,thick](3.5,0.7*1.5 + 0.5) -- (4,0.7*1.5 + 0.5) ; 
    \draw [black,thick](3.5,1.8*1.5 + 0.5) -- (4,1.8*1.5 + 0.5) ; 
    \draw [black,thick](5,4.5) -- (5.5,4.5) ; 
    \draw [black,thick](2,25/28 + 4.5 ) -- (2.5,25/28 + 4.5) ; 
    \draw [black,thick](2,45/28 + 4.5) -- (2.5,45/28 + 4.5) ; 
    \draw [black,thick](5,110/28 + 4.5) -- (5.5,110/28 + 4.5) ; 
    \draw ( 5.5 + 0.15 ,110/28 + 4.5 - 0.1)  node {\tiny \bf{$(3)$}} ;
    \draw [black,thick](3.5,135/28 + 4.5) -- (4,135/28 + 4.5) ; 
    \draw ( 4 + 0.15 ,135/28 + 4.5 - 0.1)  node {\tiny \bf{$(3)$}} ;

    \draw [mycolor10]( 2.25 ,0.35*1.5 + 0.5)  node {\normalsize $\lambda_{10} = 205 \mu$m} ;
    \draw  [mycolor21]( 2.25 ,1.3*1.5 + 0.5)  node {\normalsize $\lambda_{21} = 122 \mu$m} ;

    \draw  [black]( 5.25 ,1.8*1.5 + 0.5)  node {\large $^3\mathrm{P}_2$} ;
    \draw  [black]( 5.25 ,0.7*1.5 + 0.5)  node {\large $^3\mathrm{P}_1$} ;
    \draw  [black]( 5.25 ,0.5)  node {\large $^3\mathrm{P}_0$} ;

\end{tikzpicture}
\caption{Grotrian diagram of the first twelve energy levels of N$^{+}$, retrieved from the \href{https://www.nist.gov/pml/atomic-spectra-database}{NIST Atomic Spectra Database}. Level energies are shown as a function of spin multiplicity. The dashed gray line marks an energy gap between 200 and 20\,000~K introduced on the $y$-axis to highlight the fine-structure levels of the ground state. Solid arrows correspond to electric-dipole allowed transitions, whereas the dashed arrow corresponds to a forbidden transition. Arrows are color-coded by the spontaneous Einstein coefficient $A_{ij}$. The label {\tiny(3)} indicates that the corresponding LS state is split into three fine-structure levels. The wavelengths of the two fine-structure lines of the ground electronic state are indicated in blue.}
\label{fig:grotrian_diagram}
\end{figure}
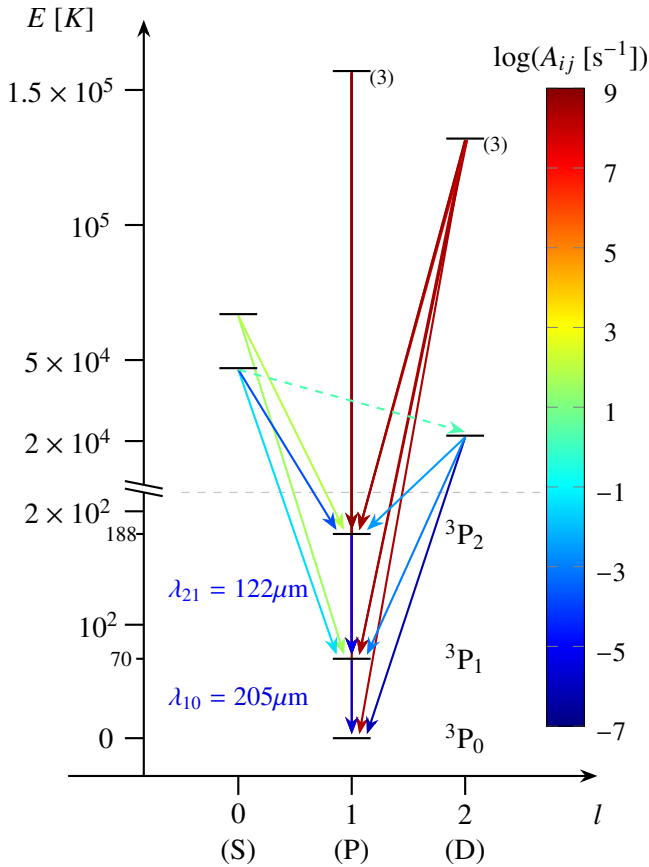 

\begin{figure*}[h!]
\centering
\includegraphics[width = 0.95\linewidth ]{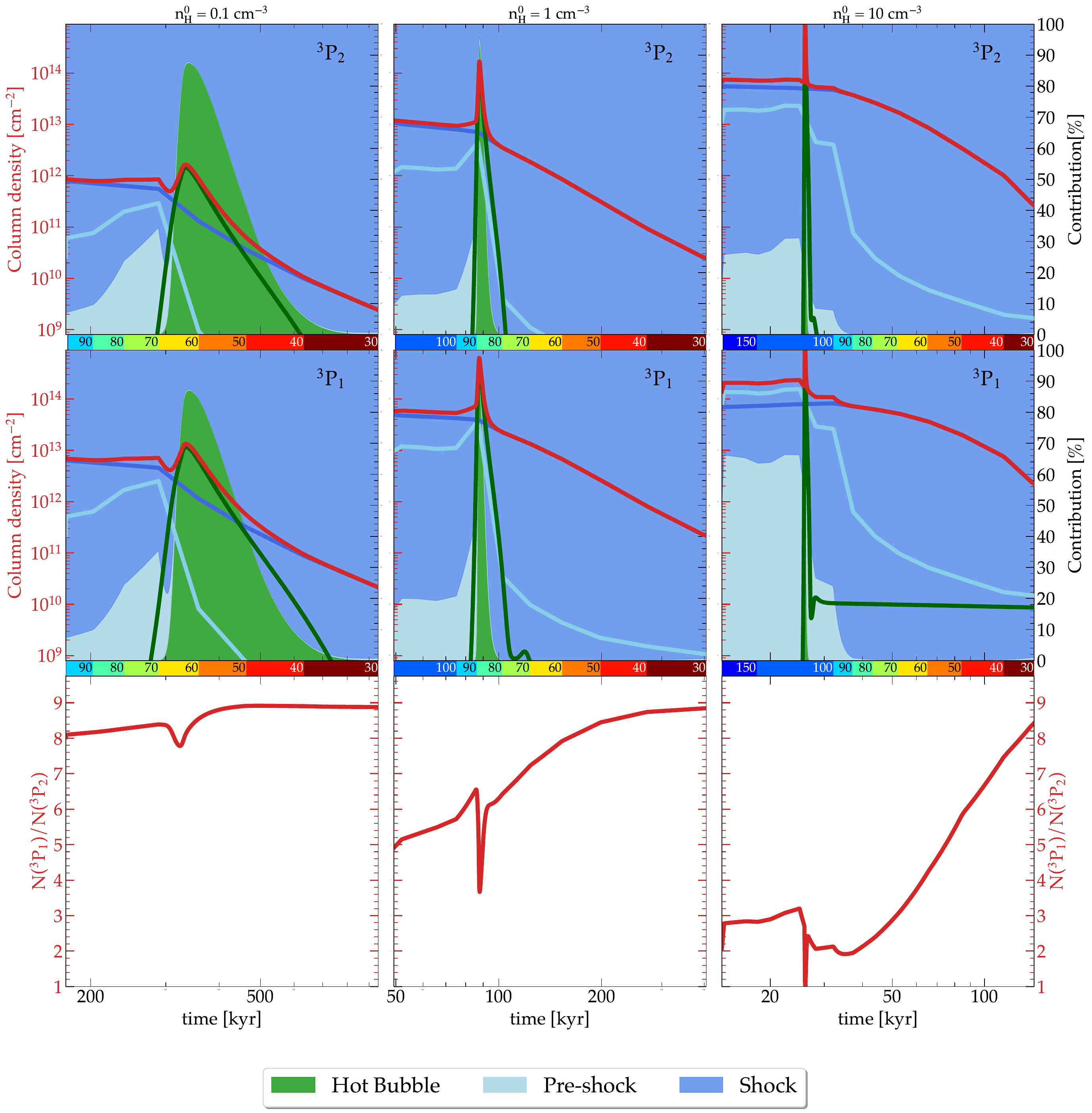}
\caption{Temporal evolution of the column densities of the $^3$P$_1$ and $^3$P$_2$ fine-structure levels of N$^+$ produced by r-SNRs and the relative contributions of the hot bubble, shocked gas, and pre-shock medium. Predictions are shown for three SNRs expanding into ambient media with $\densini = 0.1$ (left panels), 1 (middle panels), and 10~\cc\ (right panels). In each case, $G_0$ and $\zeta_{\HH}$ scale linearly with $\densini$, while $B_\perp$ scales as the square root of $\densini$ (see Sect.~\ref{sec:phys-cond}). The column densities of the $^3$P$_1$ (middle panels), $^3$P$_2$ (top panels) levels and their ratio (bottom panels) are shown as red curves and are computed along a radial line of sight crossing the spherical shell once. The fractional contributions of the hot bubble, shocked gas, and pre-shock medium are indicated in green, dark blue, and light blue, respectively.}
\label{fig:partnum_n+}
\end{figure*}

The production of N$^+$ in r-SNRs is investigated using a grid of individual r-SNR models sampled at different evolutionary stages. Because the Galactocentric scalings adopted in Eqs.~\ref{eq:GC_G0}--\ref{eq:GC_Bperp} tie together the physical parameters of the ambient medium, the dimensionality of the parameter space is reduced. The grid is therefore constructed by sampling only the ambient proton density, which in turn sets the other environmental parameters. To cover the full range of ambient proton densities explored by the fiducial Galactic model (see Table~\ref{tab:param_galac}), the grid homogeneously samples \densini\ on a logarithmic scale from $10^{-2}$ to $10^{2}$~\cc, using 40 discrete values. The temporal evolution of the r-SNRs is explored by sampling the terminal shock velocity in steps of $10$~\kms, which provides an unequivocal proxy for the evolutionary time (see Appendix~\ref{app:phase_SNR}).

The energetic structure of singly ionized nitrogen is displayed in Fig.~\ref{fig:grotrian_diagram}, which shows the first twelve levels of N$^+$ along with their associated radiative transitions. The excitation of N$^+$ in its fine-structure states within the hot bubble, shock, and pre-shock regions is computed considering collisional excitation by electrons, radiative pumping of electronic states by the ambient UV field followed by fluorescence, and chemical excitation at formation. Across the full grid of models, the excitation is found to be largely dominated by inelastic, non-reactive collisional processes.

Predictions from the individual r-SNR models are shown in Fig.~\ref{fig:partnum_n+}, which presents the temporal evolution of the radial column densities of the $^3$P$_1$ and $^3$P$_2$ fine-structure levels of N$^+$ for SNRs expanding into ambient media with different physical conditions. Several microphysical pathways contribute to the production of ionized nitrogen: (i) the successive recombination of multiply ionized nitrogen during the cooling of the hot bubble and the cooling of the post-shock gas for high terminal shock velocities ($V_B \gtrsim 100$~\kms); (ii) direct collisional ionization of neutral nitrogen in the post-shock gas for terminal shocks at lower velocity; and (iii) photoionization processes in the pre-shock medium.

Figure~\ref{fig:partnum_n+} reveals several fundamental trends. Depending on the physical conditions of the ambient medium, each of the three components of the remnant—the hot bubble, the shocked gas, and the pre-shock medium—may dominate the production of N$^+$ at different stages of the evolution and must therefore be taken into account. Although r-SNRs expanding into lower-density media persist for longer times, they are globally less efficient at producing N$^+$. In each model, N$^+$ is produced predominantly during the earliest phases of the r-SNR evolution, either during the cooling of the hot bubble or while the terminal shock velocity remains above $\sim70$~\kms. Finally, the column densities of the two fine-structure levels follow the same temporal evolution, and their ratio varies by less than a factor of five across all the models shown in Fig.~\ref{fig:partnum_n+}. This behavior indicates that, regardless of which region dominates the production of N$^+$ at a given time, the gas contributing most to [N\,II] emission is characterized by a remarkably narrow range of physical conditions.


\subsection{Methodology}

\begin{figure}[t]
\centering
\includegraphics[width = \linewidth ]{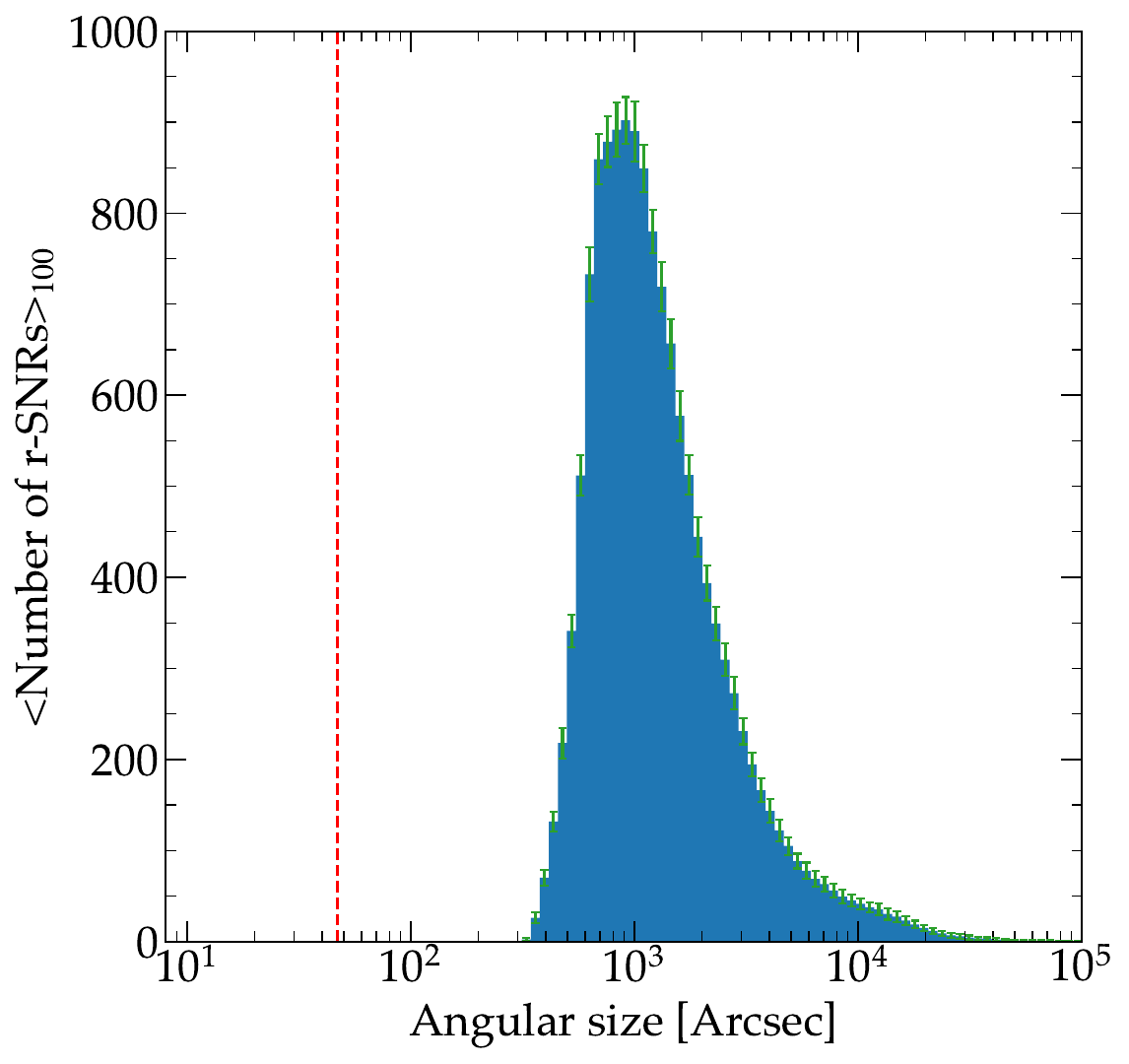}
\caption{Histogram of the angular diameters of r-SNRs, as seen from Earth, predicted by the Galactic model. The histogram (in blue) shows the mean distribution computed with 100 realizations using the fiducial parameters listed in Table~\ref{tab:param_galac}. The green error bars indicate the standard deviation in each angular-diameter bin. The red dashed line shows the PACS field of view for comparison.}
\label{fig:angular_size}
\end{figure}

In the following, the PACS data are compared quantitatively with the model predictions, while the HIFI data and the associated velocity information are treated at a qualitative level. The PACS instrument consists of 25 spaxels covering a total field of view of $47''\times47''$. Figure~\ref{fig:angular_size} compares this field of view to the distribution of angular diameters of r-SNRs, as seen from the Earth, predicted by the Galactic model. Regardless of their position in the Galaxy, all r-SNRs have angular diameters at least about an order of magnitude larger than the observational field of view. The PACS measurements along individual lines of sight can therefore be considered as pencil-beam across the spherical r-SNR structures, without any beam dilution.

Random realizations of the distribution of r-SNR population are drawn from the Galactic model. Because it is impractical to compute a dedicated model for each r-SNR in a given realization, the thermochemical structure of each remnant—including the hot bubble, shock, and pre-shock regions—is reconstructed by selecting the closest model in the precomputed grid described in Sect.~\ref{sec:prodN+_SNR} which spans a wide range of ambient conditions and evolutionary stages.

To account for the radial variation of elemental nitrogen abundance in the Galaxy, we adopt a Galactocentric gradient \citep{Arellano2020} of 
\begin{equation}
\frac{d[{\rm N}]}{dR} = -6\times10^{-2}\ {\rm dex}\ {\rm kpc}^{-1},
\end{equation}
where [N] is the elemental abundance relative to hydrogen. This gradient leads to an increase of [N] by a factor of $\sim 2$ at a Galactocentric radius of 2~kpc relative to the Solar value (see Table~\ref{tab:elem}). As nitrogen remains a trace element, this variation does not affect the thermochemical structure of the r-SNR models and is therefore applied as a simple scaling factor to the density profiles of N$^+$ and its excited levels.

The total column densities of N$^+$ and of its three fine-structure levels are computed along lines of sight across the Galactic plane on a regular grid of Galactic longitudes spaced by $1^\circ$. Under the pencil-beam approximation, the column densities along a given line of sight are obtained by summing the contributions from all r-SNR shells intersected by the ray, taking into account the geometric path length through each shell. The resulting angular distributions, which include absolute column densities, their ratios, and the associated dispersions, are then compared with the observational sample. The N$^+$ column densities are found to remain systematically below $10^{17}$~cm$^{-2}$, providing a model-based justification for applying the optically thin approximation \citep{Goldsmith2015}.


\subsection{Comparison with observations}
\label{sec:comp_with_obs}

\begin{figure}[t!]
\centering
\includegraphics[width = \linewidth ]{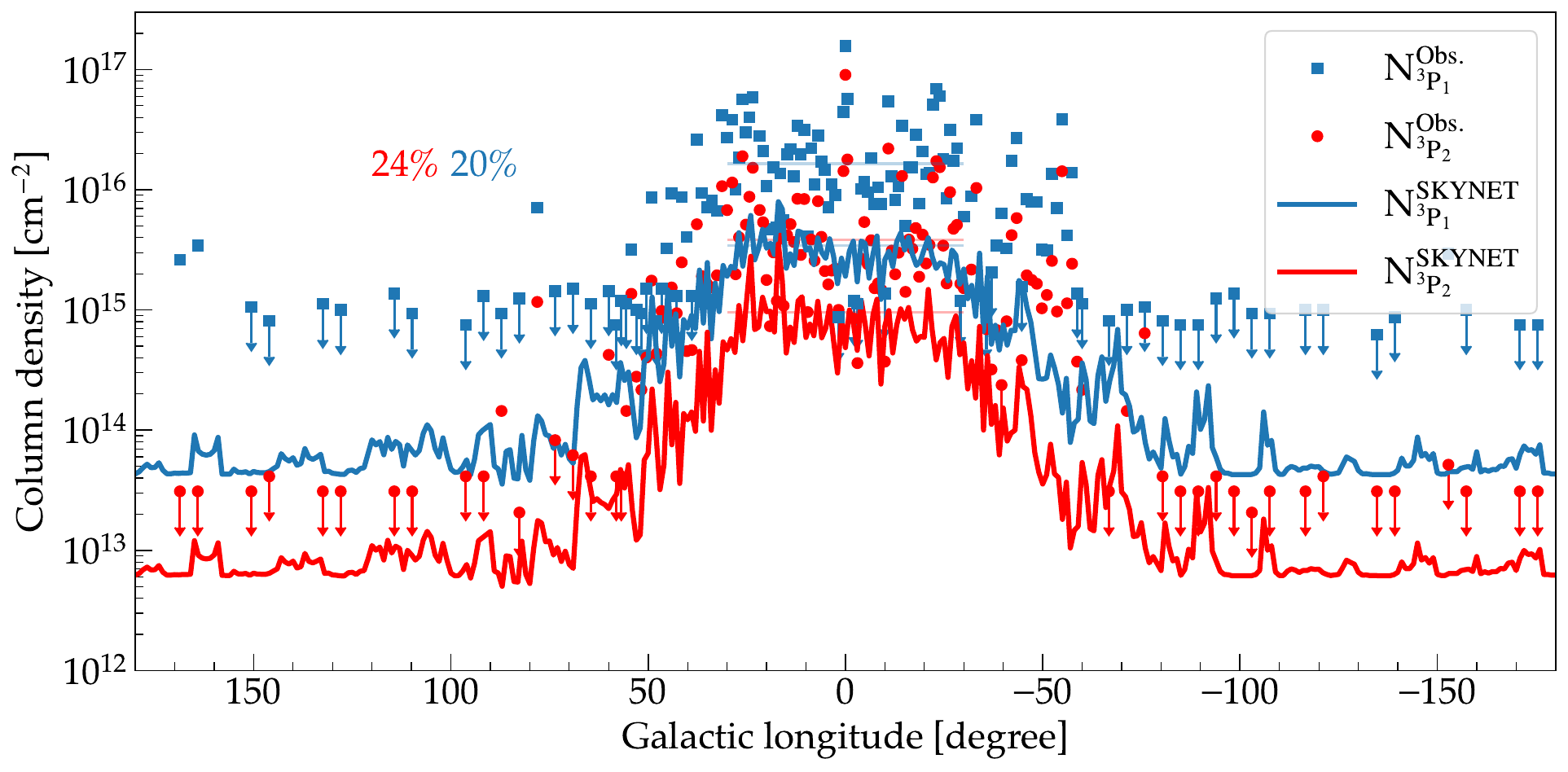}
\caption{Comparison of the observed and modeled column densities of the $^3$P$_1$ and $^3$P$_2$ levels of N$^+$ along the Galactic plane. The solid lines show the predictions from one Galactic realization generated with SKYNET. The points correspond to the column densities derived from PACS observations. The column densities of the $^3$P$_1$ and $^3$P$_2$ levels are shown in blue and red, respectively. The labels indicate the percentage of the observed column density of each level in the inner Galaxy ($|l|<30^\circ$) reproduced by the model.}
\label{fig:coldens_n+}
\end{figure}

\begin{figure}[t!]
\centering
\includegraphics[width = \linewidth ]{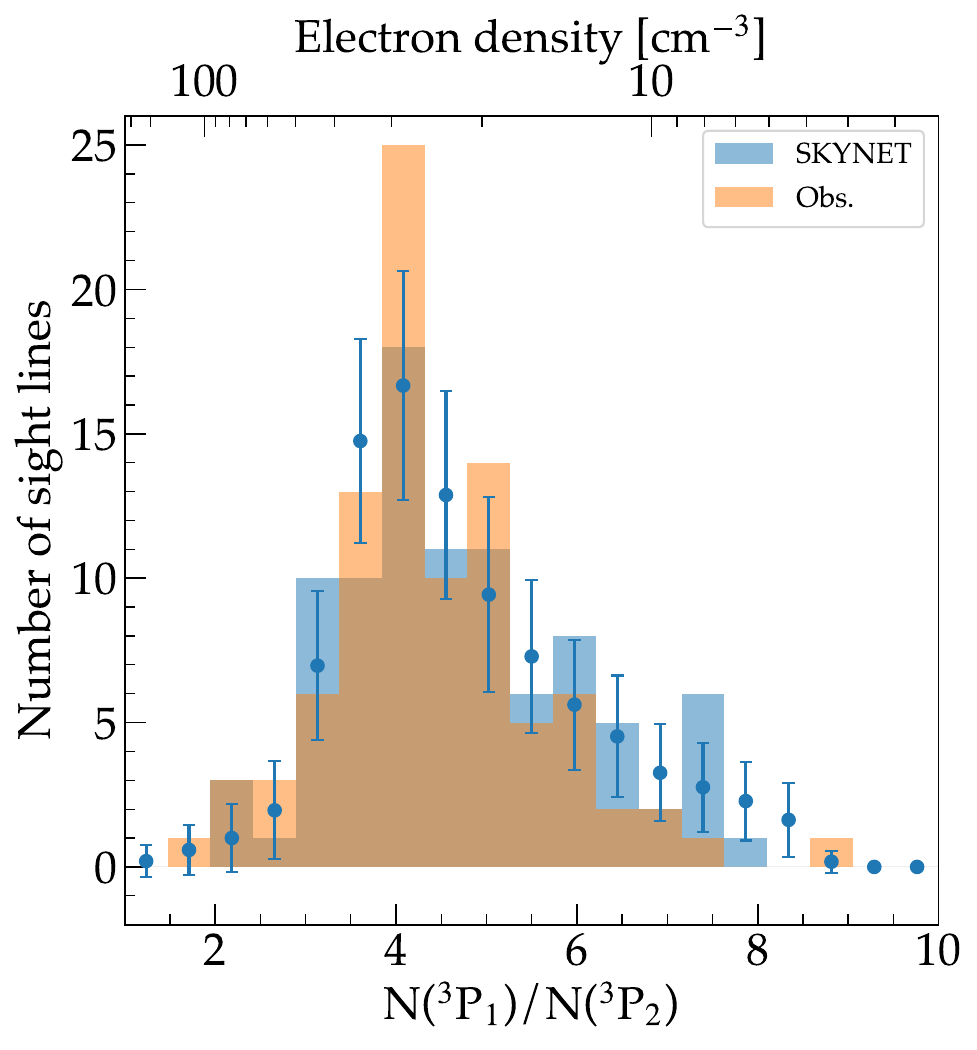}
\caption{Histogram of the N$(^3{\rm P}_1)$/N$(^3{\rm P}_2)$ column-density ratio of N$^+$ predicted by SKYNET for one Galactic realization (blue) and derived from the observations (orange). The blue points indicate the mean values obtained by averaging the model predictions over 100 Galactic realizations, while the blue error bars show the statistical uncertainty in each bin. The upper x-axis indicates the electron density corresponding to the column-density ratio for a homogeneous gas at a kinetic temperature of 8000~K \citep{Goldsmith2015}.}
\label{fig:ratio_coldens_n+}
\end{figure}

Comparisons between SKYNET predictions and the observations are shown in Fig.~\ref{fig:coldens_n+}, which displays the longitudinal profiles of the column densities of the two fine-structure excited levels of N$^+$, and in Fig.~\ref{fig:ratio_coldens_n+}, which shows the distributions of the corresponding column-density ratios along all modeled and observed lines of sight. These comparisons reveal three main results.

First, SKYNET successfully reproduces the main morphological features of the Galactic [N\,II] emission. As observed, SKYNET predicts a rise in the N$^+$ column density toward the inner Galaxy around $l = \pm 60^\circ $, corresponding to the tangential interception of the Sagittarius spiral arm (see Fig.~\ref{fig:SNR_galax_distrib}). This rise is asymmetric, with a steeper increase at positive longitudes, in agreement with the observations. Within the inner Galaxy ($-30^{\circ} \leq l \leq +30^{\circ}$), both the observations and the model exhibit a plateau in the column density with significant variations. At large longitudes ($|l| \gtrsim 70^{\circ}$), the predicted column densities are approximately two orders of magnitude lower than those in the inner-Galaxy plateau, consistent with the upper limits measured in the observational sample. One notable feature not reproduced by the model is the enhancement of [N\,II] emission around $l\sim80^\circ$, which may be associated with the Local Orion Arm, currently not included in the model.

Second, Fig.~\ref{fig:ratio_coldens_n+} reveals an excellent statistical agreement between the observed and modeled distributions of the N$(^3$P$_1)/$N$(^3$P$_2)$ column density ratio. SKYNET accurately reproduces the mean value and the width of the distribution, while also recovering the observed asymmetry toward high ratios. This agreement demonstrates that the model naturally explains the unexpectedly narrow range of physical conditions in the medium responsible for the observed [N\,II] emission. The slight excess of modeled ratios above 6 (15 lines of sight in the model, compared with 6 in the observations) is likely the result of an observational selection effect. At such large ratios, one of the two lines becomes too faint to be detected, excluding 26 lines of sight (17\% of the total sample) from the observational distribution.

Last, Fig.~\ref{fig:coldens_n+} shows that SKYNET accounts for approximately 20-25\% of the total Galactic N$^+$ column density. This result demonstrates that r-SNRs are a substantial contributor to the Galactic N$^+$ reservoir and, therefore, to the ionization of the diffuse Galactic ISM. It highlights a role that has largely been overlooked in previous studies, which generally attribute the ionization of the diffuse Galactic ISM primarily to H\,II regions. The difference between the predicted and observed N$^+$ content may reflect either simplifying assumptions in the current implementation of SKYNET or additional sources of ionization unrelated to SNRs. The origin of the remaining N$^+$ content is discussed in Sect.~\ref{discus:origin}, where we examine both possibilities.

Taken together, all these results demonstrate that SKYNET captures the key morphological and excitation properties of the Galactic N$^+$ emission and reveal that r-SNR contribute significantly to the ionization of the Milky Way.

\subsection{Origin of the modeled N$^+$ emission}
\label{sec:modelN+_origin}

\begin{figure}[t!]
\centering
\includegraphics[width = \linewidth ]{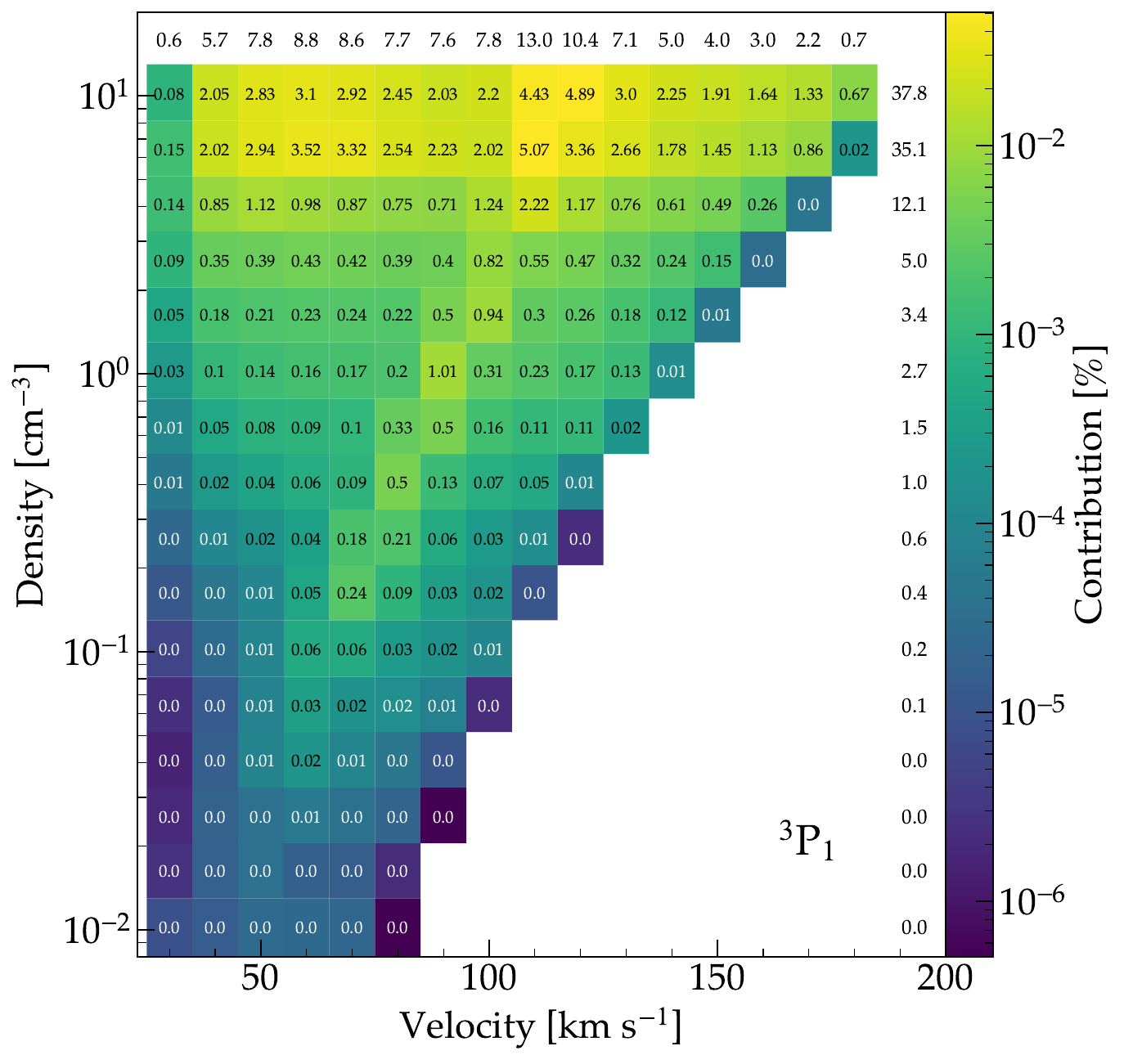}
\caption{Fractional contribution to the total predicted Galactic column density of the $^3$P$_1$ level of  N$^+$  for $|l| \leq 30 ^{\circ}$. Each cell displays the relative contribution (in percent) of r-SNRs as a function of terminal shock velocity and ambient density. The top row and rightmost column give the marginal sums over all ambient densities and all shock velocities, respectively.}
\label{fig:origin_velo_dens}
\end{figure}

The  N$^+$ emission predicted by SKYNET results from the collective contribution of r-SNRs spanning a wide range of evolutionary stages and ambient conditions. Figure \ref{fig:origin_velo_dens} shows the fractional contribution of each ($n^0_{\mathrm{H}}$, $V_{\mathrm{B}}$) cell to the total N$^+$ $^3$P$_1$ column density integrated over the inner-Galaxy plateau ($|l| \leq 30 ^{\circ}$; see Fig. \ref{fig:coldens_n+}).

\begin{figure}[t!]
\centering
\includegraphics[width = \linewidth ]{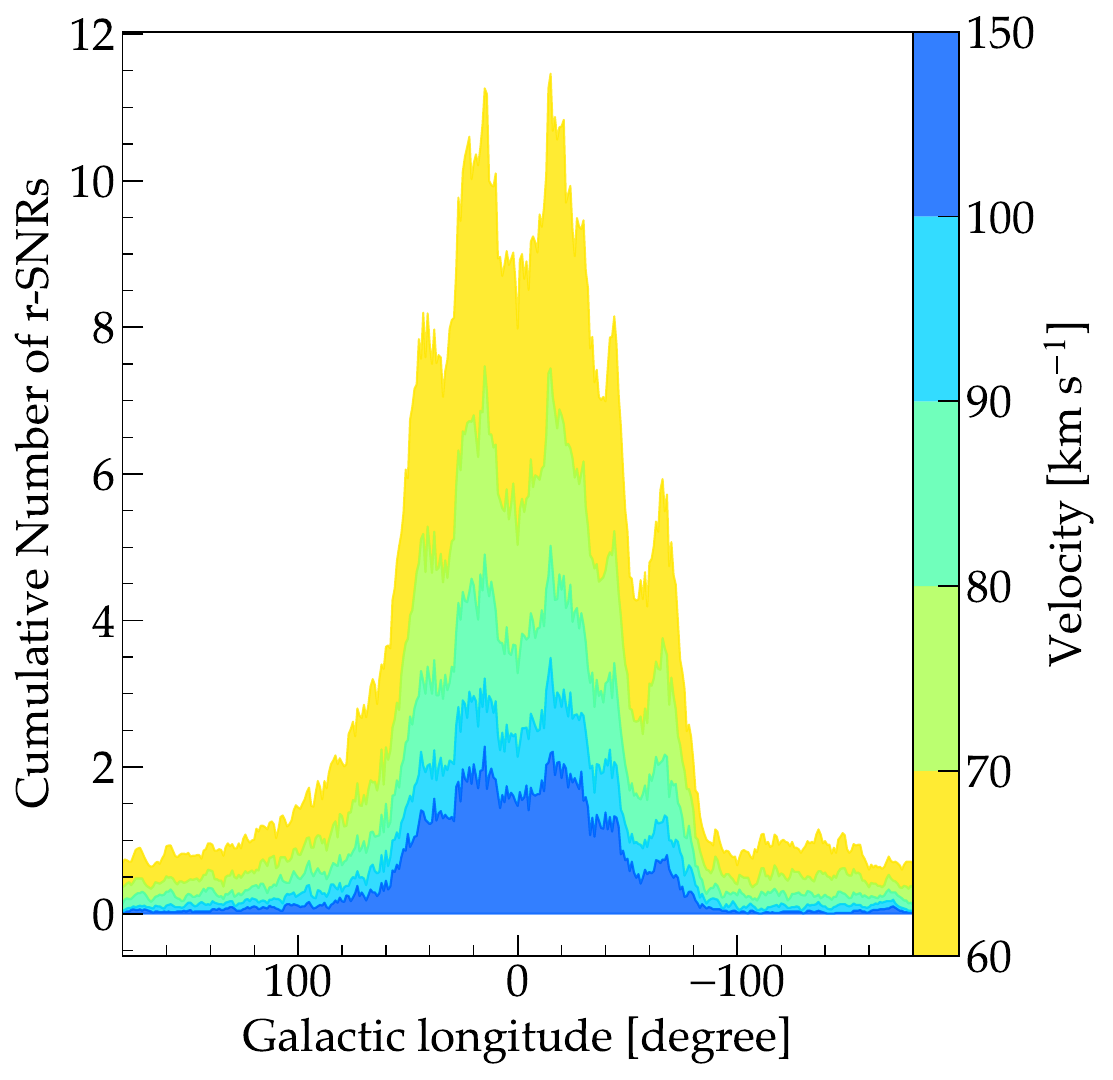}
\caption{Mean number of r-SNR shells intersected per pencil-beam line of sight at zero Galactic latitude, as a function of Galactic longitude. These values are averaged over 100 Galactic realizations. The stacked histogram shows the decomposition by terminal shock velocity. For clarity, only r-SNRs with terminal shock velocities above 60~\kms\ and below 150~\kms\ are shown.}
\label{fig:intersection}
\end{figure}

In terms of shock velocity, remnants with $V_{\mathrm{B}} <  60$~\kms\ contribute less than 15\% of the total predicted N$^+$ column density, even though they represent approximately 70\% of all modeled r-SNRs and, by virtue of their larger radii, are geometrically the most likely to intercept a given line of sight. At the other extreme, the fastest shocks contribute less than 10\% at any ambient density. The N$^+$ emission is therefore dominated by r-SNRs with intermediate terminal shock velocities. As an illustration, Fig.~\ref{fig:intersection} shows the cumulative number of r-SNR shells intercepted along lines of sight through the Galactic plane. In the longitude range where the N$^+$ column density reaches a plateau, the lines of sight intercept, on average, 4 to 12 intermediate-velocity shells. This is broadly consistent with, although somewhat higher than, the small number of velocity components associated with the observed N$^+$ emission in the HIFI data. The excess may result from the clustering of supernovae, as discussed in Sect.~\ref{discus:origin}.

The contribution of r-SNRs also depends strongly on the ambient density. Since the fine-structure lines of N$^+$ are treated as optically thin, lines of sight directed toward the inner Galaxy integrate contributions from all modeled ambient densities across the full depth of the Galactic disk. Nevertheless, r-SNRs that expand into environments with densities greater than 3 cm$^{-3}$ account for more than 80\% of the predicted column density. This dominance arises because remnants expanding into dense environments are both more numerous (see Fig.~\ref{fig:intersection}) and individually more efficient at producing N$^+$ (see Fig.~\ref{fig:partnum_n+}). These two effects more than compensate for their lower probability of being intercepted along a given line of sight.

A diagonal feature is also visible in Fig.~\ref{fig:origin_velo_dens}. This feature corresponds to the contribution of the hot bubble to the Galactic N$^+$ reservoir, which accounts for approximately 10-20\% of the total N$^+$ column density predicted by the model.

All these findings provide a coherent physical interpretation of the predicted line ratio distribution shown in Fig. \ref{fig:ratio_coldens_n+}. The mean ratio of $\sim 4$ arises naturally from the dominance of r-SNRs expanding in dense environments, which produce column density ratios between 2 and 6 (see Fig.~\ref{fig:partnum_n+}), and from the contribution of the hot bubble, which produces similar ratios at intermediate densities. Interestingly, no single combination of shock velocity and ambient density dominates the N$^+$ emission budget. Instead, the observed distribution results from the cumulative contribution of remnants spanning a broad range of physical conditions. This averaging over a broad range of remnant properties naturally makes the predicted line ratio largely insensitive to the precise value of any individual model parameter. The robustness of this result is quantified in the following section.

\subsection{Parameter dependence}

\begin{figure*}[h!]
\centering
\includegraphics[width = \linewidth ]{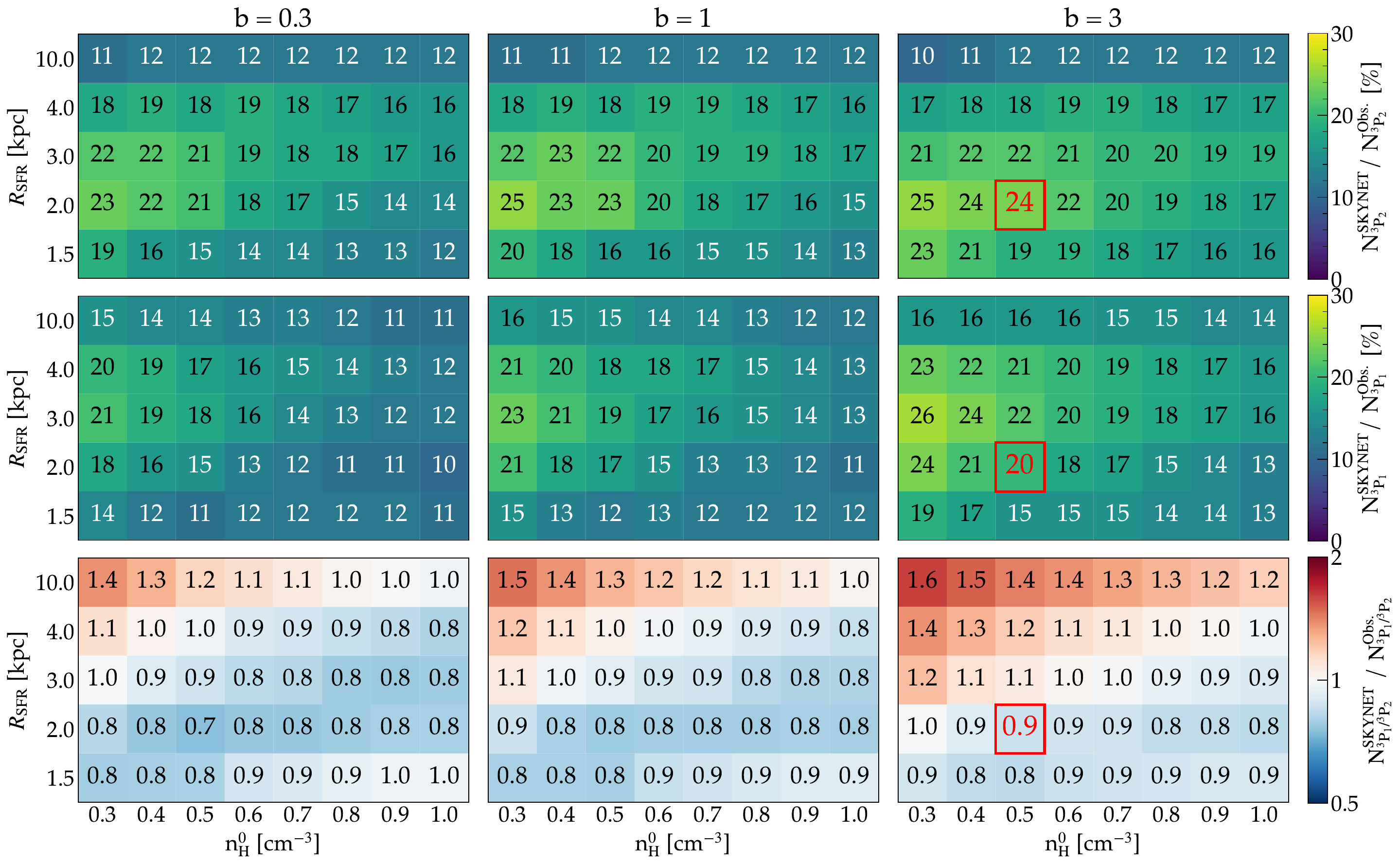}
\caption{Comparisons of SKYNET predictions with the observations as functions of the main parameters. Top (resp. middle) panels display the ratio (in percent) of the model-predicted to observed column density of the $^3{\rm P}_1$ (resp. $^3{\rm P}_2$) level of N$^+$ in the inner Galaxy ($|l|<30^\circ$). The bottom panels show the ratio of the median SKYNET-predicted to the median observed N$(^3$P$_1$)/N($^3$P$_2$) column-density ratio. All ratios are shown as functions of the solar total proton density $n_\mathrm{H}^0$, the star-formation-rate scale radius $R_\mathrm{SFR}$, and for three values of the magnetic field parameter $b$ ($b = 0.3$, $1$, and $3$ from left to right). The cell outlined in red identifies the standard model and its associated parameters (see Table~\ref{tab:param_galac}).}
\label{fig:dep_param}
\end{figure*}

All the results presented so far have been obtained with the standard Galactic model defined by the parameters listed in Table~\ref{tab:param_galac}. As shown in Table~\ref{tab:param_galac}, SKYNET involves approximately 20 independent parameters. At first sight, such a large number of parameters may suggest that SKYNET is a highly flexible model with limited predictive capability. This is not the case in practice. As discussed in Sect.~\ref{sec:galac_distribution}, most parameters are independently constrained by observations, each with only moderate uncertainties. Furthermore, many parameters either have a negligible influence on the predicted N$^+$ emission or affect the model results in a simple, analytically predictable manner. 

As shown in Appendix~\ref{app:phase_SNR}, the ejected mass $M_{\rm ej}$ has no influence on the model predictions during the radiative stage because the swept-up mass largely exceeds $M_{\rm ej}$ before this phase begins. Similarly, the onset scaling factors of the PD and MC stages, $\alpha$ and $\beta$, have a negligible impact. On the one hand, the shocks responsible for most of the N$^+$ emission occur well below the maximum shock velocities (see Sect.~\ref{sec:modelN+_origin}), making the results insensitive to $\alpha$. On the other hand, the PD and MC phases follow nearly identical power-law evolutions, making the predictions insensitive to $\beta$. Finally, because the cosmic-ray and UV energy fluxes are negligible compared with the mechanical energy flux carried by radiative shocks, the corresponding parameters, $\zeta_{\HH}^{\odot}$ and $G_0^{\odot}$, do not significantly affect the shock thermodynamics and therefore have a negligible influence on the predicted emission.

Among all parameters, the supernova rate $k_{\rm SN}$ and the HIM filling factor $\phi$ have analytically predictable effects on the model predictions. Increasing $k_{\rm SN}$ simply increases the number of r-SNRs intercepted along a given line of sight without modifying the physical conditions within individual remnants. Consequently, the predicted column densities scale linearly with $k_{\rm SN}$, while the N$^+$ column-density ratio remains unchanged. The HIM filling factor $\phi$ affects the predictions only through the size of radiative remnants, hence their probability of interception along a given line of sight. Varying $\phi$ over its estimated range of 0.25-0.75 changes the predicted column densities by less than 30\%, while leaving the column-density ratio essentially unaffected.

Only three parameters may substantially and non-trivially affect the predicted N$^+$ emission: the solar-neighborhood proton density $n_\mathrm{H}^0$, the star-formation-rate scale radius $R_\mathrm{SFR}$, and the magnetic-field parameter $b$. The first two determine the normalization and radial profile of the ambient density, while $b$ controls the resistance of the shocked gas to magnetic compression. The influence of these three parameters is summarized in Fig.~\ref{fig:dep_param}, which compares the SKYNET predictions with the observations over the $(n_\mathrm{H}^0,R_\mathrm{SFR},b)$ parameter space. The top and middle panels show the predicted-to-observed column-density ratios for the two fine-structure levels of N$^+$, while the bottom panels display the predicted-to-observed median N$(^3$P$_1$)/N$(^3$P$_2)$ column-density ratio. This figure reveals that the SKYNET predictions depend only weakly on these parameters: over the wide range of values explored, both the predicted column densities and the column-density ratio vary by at most a factor of about two.

The magnetic-field parameter $b$ has the weakest influence. For fixed values of $R_\mathrm{SFR}$ and $n_\mathrm{H}^0$, varying $b$ changes neither the predicted column densities nor the column-density ratio by more than a few percent. This confirms that, within the range explored, the magnetic field strength plays only a minor role in shaping the Galactic N$^+$ emission. The radial density profile has a more noticeable effect. Both very steep ($R_\mathrm{SFR}=1.5$ kpc) and very flat ($R_\mathrm{SFR}=10$ kpc) density profiles produce lower column densities than the standard model. Flat profiles additionally yield higher column-density ratios because they reduce the contribution of the dense inner Galaxy, where the N$^+$ emission is strongest and the column-density ratio is lowest (see Figs.~\ref{fig:origin_velo_dens} \& \ref{fig:partnum_n+}). In most cases, decreasing $n_\mathrm{H}^0$ increases both the predicted column-density ratio and the absolute column densities. The former follows the same trend as in the single-remnant predictions shown in Fig.~\ref{fig:partnum_n+}, as the ratio is essentially unaffected by the probabilistic distribution of SNRs. By contrast, the trend in the absolute column densities differs from the single-remnant case because they also depend on the probability that random lines of sight intercept SNRs. The only exception to the trend in the column-density ratio occurs for the steepest density profile, where the increasing contribution of high-velocity shocks, dominated by emission from the pre-shock gas, produces a slight increase in the ratio with $n_\mathrm{H}^0$.

Taken together, these results demonstrate that SKYNET predictions effectively depend on only a handful of physically relevant parameters. More importantly, the exploration of the parameter space reveals that the model provides remarkably robust predictions for both the absolute column densities of the N$^+$ fine-structure levels and their column-density ratio. Across the range of parameters explored in this work, both observables vary only weakly, highlighting that the predicted Galactic N$^+$ emission originating from r-SNRs is an intrinsic consequence of the underlying physics rather than the result of a particular choice of model parameters.

\section{Discussions}
\label{sec:discussion}

\subsection{Current scope of SKYNET}

Despite its simplicity, SKYNET already yields several results that are independent of the detailed model assumptions. First, and as already emphasized by \citet{Godard2024b}, r-SNRs are unavoidable along lines of sight across the Galactic plane and leave a detectable cumulative imprint on specific tracers such as the fine-structure lines of [C\,I] and [N\,II]. Comparisons with large-scale Galactic surveys therefore offer an opportunity to investigate the statistical properties of r-SNR populations. Second, the spatial distribution of r-SNRs naturally produces large-scale longitude variations in emission tracers that reflect the underlying Galactic structure. These trends arise not only from dynamical features such as spiral arms, but also from systematic variations in the local physical conditions, including density, ionization rate, and magnetic field strength. These conditions are themselves closely connected to the distribution of the diffuse atomic and molecular gas, traced by HI and CO, and ultimately set by the gravitational potential of the Galaxy, thermal instability, and interstellar turbulence. Third, different tracers probe different phases of the r-SNR evolution and therefore sample distinct thermodynamic regimes. For example, [C\,I] emission has been shown to trace the wide range of thermal pressures generated by SNR-driven shocks \citep{Jenkins2011,Godard2024b}, while [N\,II] emission arises from a remarkably narrow excitation regime in terms of electron density and temperature.

This being said, the SKYNET framework results from the combination of several independent idealized models and relies on several simplifying hypothesis. Although the results obtained for [N\,II] emission are found to be relatively insensitive to many of these assumptions and parameters, other tracers may be more sensitive to the choices adopted here. In addition, not all assumptions and parameters have yet been explored systematically. In its current form, the SKYNET framework should therefore be regarded as an initial implementation of a model of a Galactic distribution of r-SNRs. It establishes the conceptual and methodological basis of the approach, assesses its predictive power, and identifies the key physical ingredients, as well as the limitations and improvements that should be incorporated in future developments.


\subsection{Limitations of the single-remnant model}
\label{sec:lim-single}

One of the main limitations of the approach stems from the one-dimensional geometry adopted for modeling individual r-SNRs. In this configuration, the shocked gas cannot escape in the azimuthal and poloidal directions around the expanding shell and is therefore artificially confined along the radial direction. As a consequence, the post-shock trajectory eventually becomes unrealistic when the built up of magnetic pressure produces what can be viewed as a magnetic wall \citep{Godard2024a}. In addition, 1D models inherently suppress the development of dynamical and thermodynamical instabilities and the fragmentation of the gas that occurs during the cooling of the hot bubble and the post-shock \citep[e.g.][]{Innes1987, Blondin1998, Falle2020, Raymond2020, Markwick2021}. Such instabilities may affect the column density structure of the r-SNR and the line emissivities. One possible solution would be to adopt different equations of state for the cooling of the hot bubble and post-shock gas and set their probability of occurrence based on the results of 3D simulations of SNRs and cooling layers. 

A related issue is the assumption that r-SNRs expand only within the most diffuse phases of the ISM, namely the WNM and the HIM, hence the absence of interactions with dense clouds. Such interactions are expected to be common in the multiphase ISM, where supernova remnants expand through a medium structured by thermal instability and turbulence \citep[e.g.,][]{Inoue2012}. Encounters with dense clumps modify the swept-up mass, internal structure, and energy budget of the remnant \citep{Guo2025}, drive shocks into dense molecular clouds, can trigger dynamical instabilities and generate turbulent mixing layers at cloud interfaces \citep[e.g.][]{Kwak2010, Kupilas2021, Makarenko2023}. All these processes may affect the resulting emission tracers and, in particular, optical line diagnostics  \cite[e.g.,][]{Slavin1993b, Makarenko2023}.  An interesting avenue to alleviate this oversight would be to divide individual r-SNRs, at a given evolutionary stage, into angular sectors interacting with different ambient media, characterized by a probability distribution function of the surrounding density, as proposed by \citet{Haid2016}.

Another limitation concerns the treatment of radiative transfer and the interaction of the ionizing photons emitted by the post-shock gas and the hot bubble with the entire r-SNR structure. Because the fully ionized hot bubble is optically thin and the gas is expected to fragment in cooling layers, we assume that these photons interact only with the pre-shock material. Such an approximation may lead to an overestimation of the ionizing flux reaching the pre-shock region. Since the contribution of the pre-shock to the total [N\,II] emission remains subdominant, this simplification is expected to have little impact on the predicted N$^+$ column densities and line intensities. This may not be the case, however, for other tracers. A more realistic treatment would require a statistical description of the fraction of photons interacting with the fragmented post-shock structure, which remains beyond the scope of the present implementation of SKYNET.

The survival of dust grains in SNRs have been the subject of numerous studies with no clear consensus yet (\citealt{Scheffler2026} and references therein). In its current implementation, the Paris–Durham shock code solely includes erosion of large grains \citep{Godard2019}. In particular, the code does not account for the fragmentation and destruction of grains by shattering and vaporization processes \citep[e.g.,][]{Jones1996, Slavin2015}, nor for the destruction of PAHs through sputtering \citep{Micelotta2010a, Micelotta2010b}. This omission may be important, as vaporization, shattering, and sputtering modify the grain size distribution, can lead to the complete destruction of PAHs in fast shocks above $\sim100$ km\,s$^{-1}$ \citep[e.g.,][]{Micelotta2010b}, and therefore affect the absorption properties of dust, their charge distribution, and the recombination rates of ionized species. Considering that PAHs are expected to be destroyed in fast shocks, we adopt here a relatively low PAH abundance relative to hydrogen of $10^{-8}$ (see Table~\ref{tab:elem}). To explore the effect on the production of ionized species, an additional grid was run with a larger PAH abundance of $10^{-6}$. As found by \citet{Godard2024b} for the production and excitation of C\,I, we find that this choice has a negligible impact on the predicted [N\,II] emission. This result is due the fact that the recombination processes leading to the formation of N$^+$ and C are dominated by radiative and dielectronic recombination rather than by charge exchange with PAHs. Still, treating the evolution of dust self-consistently remains a priority for future developments. This could be achieved either by treating dust grains as particles and following their trajectories in the magnetic field structure \citep[e.g.,][]{Guillet2007, Guillet2009}, or by treating dust at different sizes and PAHs as separate fluids \citep[e.g.,][]{Slavin2024, Verrier2025} and adopting effective destruction and fragmentation rates driven by their relative velocities and velocities relative to the neutral and ionized gas.

A final aspect missing from the modelling of individual SNRs is the acceleration of particles through diffusive shock acceleration, which may lead to the energization of cosmic rays and the hardening of their spectrum even during the radiative phase of the remnant \citep{Zirakashvili2022,Cristofari2025}. Because the associated energy losses are expected to remain negligible compared to radiative losses, this process is neglected here. An elegant way to incorporate this mechanism and provide predictions for the associated synchrotron emission and other observational tracers would be to adopt a semi-analytical treatment of particle acceleration as proposed by \citet{Cristofari2025}.



\subsection{Limitations of the Galactic framework}
\label{discuss:Galframe}

Simplifications also arise in the Galactic framework. The most important one is the assumption that SNRs expand without mutual interactions. This is unlikely to hold for core-collapse supernovae, since massive stars form predominantly in clusters and OB associations \citep{deWit2005, Zinnecker2007, Tan2014}, implying that their explosions are correlated in both space and time \citep{Fielding2018, Zapartas2017}. Such clustering is expected to affect the porosity of SNRs, the volume filling factor of the HIM, the fraction of supernova energy retained in hot gas, and their radiative losses because sequential explosions can overlap, inflate superbubbles, and drive Galactic winds \citep{Tomisaka1986, Yadav2017, Fielding2018}. In the context of SKYNET, this limitation is likely to affect the strength and variance of line emission across the Galactic plane. A natural extension of the model would be to distinguish between isolated and clustered supernovae: the former could still be treated as independent remnants, while the latter could be grouped into common explosion sites and evolved as collective bubbles powered by sequential energy injection. Such an extension remains compatible with spherically symmetric modelling, as shown by previous analytical and one-dimensional studies of superbubbles driven by clustered supernova feedback \citep{Tomisaka1986, MacLow1988, Sharma2014, Orr2022}.

Another, though less critical, aspect concerns the geometrical description of the Galaxy. To reduce the number of free parameters, the Galaxy is modeled with four identical spiral arms. This simplified geometry does not account for other known structures, such as the Local arm, nor for the asymmetries between different spiral arms. In addition, the model does not include inner Galactic structures such as the Galactic bar \citep{Wegg2015, BlandHawthorn2016} and the central molecular zone. These simplifications are likely to affect the predicted spatial distribution of emission tracers \citep{Kachelriess2025} and probably explain why the longitudinal profile of [N\,II] emission across the Galactic plane does not perfectly reproduce the observations (see Fig.~\ref{fig:coldens_n+}). A more realistic description could be implemented at the cost of introducing a significantly larger number of free parameters.

To reduce the dimensionality of the problem, the density, irradiation, ionization, and magnetization of the ambient gas in which r-SNRs expand are tied together through physically motivated radial scaling laws that ensure the existence of a multiphase ISM (see Sect.~\ref{sec:phys-cond}). These scalings lead to a radial variation of the cosmic-ray ionization rate that is broadly consistent with observational constraints derived in the Solar Neighbourhood \citep[e.g.,][]{Indriolo2015,Neufeld2017} and in the Galactic Centre \citep{LePetit2016}. The resulting variation of the ambient UV radiation field is comparable, though somewhat larger (by about a factor of two), than that inferred from recent radiative transfer models of the Milky Way \citep{Popescu2017, Natale2022}. On the one hand, the scaling adopted here therefore carries significant uncertainties related to the Galactic structure, the star formation rate, the flaring of the disc, and the extent of the inner Galaxy. On the other hand, these prescriptions represent only mean environmental conditions. A more realistic approach would be to draw local values of environmental parameters from probability distribution functions informed by radiative transfer models applied to infrared observations \citep{Natale2022} and by the distribution of density obtained in numerical simulations of the turbulent multiphase ISM \citep[e.g.,][]{Audit2010, Kritsuk2017, Bellomi2020, Kobayashi2022}.

Another related issue is the absence of any explicit link between these environmental conditions and the spiral-arm structure of the Galaxy. Spiral arms are commonly interpreted as density waves that locally enhance the gas density and concentrate the star formation as they propagate through the disk \citep[e.g.,][]{Dobbs2014}. As a consequence, several key environmental parameters — including the gas density, the UV radiation field, the cosmic-ray ionization rate, and the magnetic field strength — are expected to vary systematically across spiral arms. A natural extension of SKYNET would be to couple the spatial distribution of environmental parameters to the spiral-arm structure while ensuring that their azimuthally averaged radial profiles remain consistent with the global scaling laws adopted in Sect.~\ref{sec:phys-cond}. Such an extension would allow the model to capture both the large-scale radial trends and the local variations associated with spiral structure.

 A final limiting aspect is the assumption regarding elemental abundances. All individual SNRs are modeled with the same distribution of elements without taking into account their variation with Galactocentric radius and Galactic height. These variations might impact the cooling of the gas and the dust content and therefore affect the thermochemical structure of SNRs beyond the simple scaling factor that we apply to compute N$^+$ emission. These variations and their effects on individual SNRs will be treated in future versions of SKYNET.

\subsection{Origin of the observed N$^+$ emission}
\label{discus:origin}

Although r-SNRs naturally reproduce the observed distribution of the N$^+$ line ratio and the large-scale Galactic morphology of the emission, SKYNET predicts that they account for only 20-25\% of the total observed N$^+$ column density. In addition, the model systematically underestimates the variance of the emission in the inner Galaxy and slightly overestimates the number of velocity components associated with the observed N$^+$ emission. As shown in Fig.~\ref{fig:dep_param}, the difference persists across the entire parameter space, which rules out a simple recalibration of the Galactic model. The missing N$^+$ emission may therefore reflect either simplifying assumptions in the model, leading to an underestimate of the contribution of r-SNRs, or the presence of additional ionizing sources unrelated to supernova remnants.

A first possibility is the clustering of supernova explosions into superbubbles, as discussed in Sect.~\ref{discuss:Galframe}. Compared to isolated r-SNRs, superbubbles sustain higher shock velocities over longer timescales and drive mechanical energy into larger volumes of the ISM. As shown in Fig.~\ref{fig:partnum_n+}, both effects enhance the production of N$^+$. Replacing isolated supernova remnants with a realistic clustered population could therefore increase the predicted N$^+$ column densities, increase their spatial variance, and reduce the number of velocity components, potentially alleviating all three discrepancies simultaneously.

A second possibility is the interaction of r-SNRs with pre-existing CNM structures, which are unavoidable in the multiphase ISM (Sect.~\ref{sec:lim-single}). As shown in Fig.~\ref{fig:partnum_n+}, shocks propagating into denser environments produce larger N$^+$ column densities despite their smaller physical extent. Such interactions could therefore increase the total N$^+$ content predicted by SKYNET. However, the higher post-shock densities also shift the level populations toward lower N$(^3$P$_1$)/N$(^3$P$_2)$ column-density ratios. The observed line-ratio distribution therefore places a strong constraint on the magnitude of this effect.

A final possibility is the presence of ionizing sources unrelated to SNRs. Classical H\,II regions are the most natural candidates, but the observed N$(^3$P$_1$)/N$(^3$P$_2)$ column-density ratio strongly constrains which types of H\,II regions can contribute. Ratios of 3-6 require relatively low electron densities, comparable to those found in extended and evolved H\,II regions ($n_{\rm e}\sim 1$-100 cm$^{-3}$; \citealt{Draine2011}). In contrast, compact H\,II regions, with electron densities up to $n_{\rm e}\sim10^4$ cm$^{-3}$, would drive the ratio below 2 and are therefore unlikely to dominate the missing [N\,II] emission. Extended and evolved H\,II regions therefore remain a plausible contributor to the fraction of the Galactic N$^+$ reservoir not reproduced by SKYNET.

\section{Conclusions}
\label{sec:ccl}

This paper introduces SKYNET, a new framework that couples a physical model of individual r-SNRs with a statistical description of their Galactic spatial distribution. The framework predicts the cumulative Galactic emission produced by entire populations of r-SNRs in nearly one million spectral lines. As a first application, we compare the SKYNET predictions with the spatial distribution of the fine-structure lines of N$^+$ at 122 and 205~$\mu$m observed in the \textit{Herschel} survey of \citet{Goldsmith2015}.

The Galactic distribution of SNRs and the properties of the surrounding medium are derived from the spiral structure and flaring of the Milky Way, assuming that the supernova rate, the ambient density, and the mean UV radiation field locally trace the volumetric star-formation rate. The standard model, constrained by independent observations, predicts a steady-state population of approximately 2200 non-radiative SNRs and 14\,000 r-SNRs. Their angular distribution indicates that most remnants are confined within $\sim 1^\circ$ of the Galactic plane, in qualitative agreement with recent MeerKAT observations of synchrotron emission. The large number of radiative remnants implies that random Galactic sightlines inevitably intercept several r-SNRs, whose cumulative emission produces detectable spectroscopic signatures.

The comparison with N$^+$ observations yields three main results. First, SKYNET successfully reproduces the large-scale morphology of the Galactic N$^+$ emission, including the absence of detections at high longitudes, the sharp increase around $|l|\sim60^\circ$, and the broad plateau across the inner Galaxy ($|l|<30^\circ$). This morphology emerges from the spiral structure of the Milky Way and the radial variation of the ambient density. Second, the predicted N$(^3$P$_1$)/N$(^3$P$_2)$ column-density ratio accurately reproduces the observed distribution, including its mean, dispersion, and asymmetric tail toward high ratios. This remarkable agreement demonstrates that r-SNRs provide a natural explanation for the narrow range of physical conditions responsible for N$^+$ emission. Finally, SKYNET accounts for approximately 20-25\% of the total Galactic N$^+$ column density. Taken together, these results reveal that r-SNRs are a significant contributor to the ionization of the diffuse Galactic ISM.

Despite the apparent complexity of the framework, the predicted N$^+$ emission remains unexpectedly robust. Many of the model parameters either have a negligible influence on the predicted emission or affect the results in an analytically predictable manner. Across a wide range of configurations, both the absolute column densities and the N$(^3$P$_1$)/N$(^3$P$_2)$ column-density ratio vary by less than a factor of two. This low sensitivity demonstrates that the predicted Galactic N$^+$ emission is a robust consequence of the underlying physics rather than of any particular choice of the model parameters.

The remaining difference between the predicted and observed Galactic N$^+$ column density likely results from missing physical ingredients. In particular, clustered supernova explosions and the interaction of r-SNRs with pre-existing CNM structures are both expected to increase the predicted N$^+$ column densities and their spatial variance, in closer agreement with the observations. Alternatively, part of the missing N$^+$ emission may originate from ionizing sources unrelated to SNRs. The most natural candidates are extended, evolved H\,II regions with electron densities of $n_{\rm e}\sim1$--$10^2$ cm$^{-3}$, whose excitation conditions are consistent with the observed line-ratio distribution. Incorporating these physical processes into future versions of SKYNET will provide a more complete description of the Galactic N$^+$ reservoir while further extending the predictive capabilities of the framework. SKYNET could thus participate to optimizing the scientific return of recent and future far-infrared balloon experiments, such as GUSTO (whose heterodyne receiver observed the 205~$\mu$m line with 0.2~km~s$^{-1}$ spectral resolution in 2023$-$2024) and ASTHROS (a 2.5~m telescope that will observe both lines studied here with heterodyne technology). SKYNET will also open new perspectives to interpret the data from longer-term prospects far-infrared space telescopes such as PRIMA or LETO.

The present study illustrates the potential of SKYNET. With its exceptionally broad spectral coverage, the framework is well suited to investigate atomic and molecular diagnostics spanning the wide range of thermodynamic conditions encountered throughout the evolution of supernova remnants. More generally, SKYNET provides a new tool for quantifying the impact of SNRs on the energetics, ionization, and chemistry of the Galactic interstellar medium. By identifying the collective spectroscopic signatures of r-SNRs across the Milky Way, SKYNET paves the way for the systematic statistical detection of a population that has so far remained largely invisible.

\begin{acknowledgements}
 We are grateful to the referee, Ekaterina Makarenko, for her thorough reading of the manuscript. The research leading to these results has received funding from the European Research Council, under the European Community’s Seventh framework Programme, through the Advanced Grant MIST (FP7/2017–2022, No. 742719). The grid of r-SNR models used in this work has been run on the computing cluster Totoro of the ERC MIST, administered by MesoPSL. We would also like to acknowledge the support from the Thematic Action "Physique et Chimie du Milieu Interstellaire" (PCMI) of INSU Programme National "Astro", with contributions from CNRS Physique \& CNRS Chimie, CEA, and CNES.
\end{acknowledgements}

%
\bibliographystyle{aa} 
\bibliography{main} 

\appendix
\section{Spherical SNR evolution} 
\label{app:phase_SNR}

\subsection{Expansion in a homogeneous medium}

Following the approach of \citet{Godard2024b}, the temporal evolution of the physical properties of individual supernova remnants is estimated by adopting the idealized picture of a spherical blast wave with an ejecta mass $M_{\rm ej}$ and an initial kinetic energy $E_{\rm SN}$, expanding into a homogeneous medium with proton density $n_{\rm H}^0$. In this configuration, the remnant evolves through four successive stages: the free-expansion (FE), Sedov–Taylor (ST), pressure-driven (PD), and momentum-conserving (MC) phases \citep[e.g.][]{Cioffi1988,Truelove1999,Kim2015}. The radius, $R_B$, and expansion velocity, $V_B$, of the SNR are described using the classical, piecewise analytical expressions \citep[e.g.][]{Draine2011,Vink2020},

\begingroup\makeatletter\def\f@size{7.10}\check@mathfonts
\begin{equation} \label{eq:rb}
R_B= 
\begin{dcases}
\phantom{1}3.1\,\,{\rm pc}\,\,\left(\frac{M_{\rm ej}}{M_\odot}\right)^{1/3} \,\, \,  \left(\frac{\densini}{n_0}\right)^{-1/3} \, \, \left(\frac{t}{t_{\rm ST}}\right) & \text{if } 0 < t \leqslant t_{\rm ST} \\
\phantom{1}3.1\,\,{\rm pc}\,\,\left(\frac{M_{\rm ej}}{M_\odot}\right)^{1/3} \,\, \,  \left(\frac{\densini}{n_0}\right)^{-1/3} \, \, \left(\frac{t}{t_{\rm ST}}\right)^{2/5} & \text{if } t_{\rm ST} < t \leqslant t_{\rm PD}\\
23.8\,\,{\rm pc}\,\, \left(\frac{E_{\rm SN}}{E_{51}}\right)^{0.29} \,\,  \left(\frac{\densini}{n_0}\right)^{-0.42} \alpha^{2/5}\ \left(\frac{t}{t_{\rm PD}}\right)^{2/7}  & \text{if } t_{\rm PD} < t \leqslant t_{\rm MC}\\
46.0\,\,{\rm pc}\,\,\left(\frac{E_{\rm SN}}{E_{51}}\right)^{0.29} \,\,  \left(\frac{\densini}{n_0}\right)^{-0.42} \alpha^{2/5}\ \left(\frac{\beta}{10}\right)^{2/7} \left(\frac{t}{t_{\rm MC}}\right)^{1/4} & \text{if } t_{\rm MC} < t \leqslant t_{\rm fade}
\end{dcases}
\end{equation}
\endgroup
and
\begingroup\makeatletter\def\f@size{7.10}\check@mathfonts
\begin{equation} \label{eq:vb}
V_B= 
\begin{dcases}
10\,000\,\,{\rm \kms}\,\, \left(\frac{E_{\rm SN}}{E_{51}}\right)^{1/2} \,\,  \left(\frac{M_{\rm ej}}{M_\odot}\right)^{-1/2} & \text{if } 0 < t \leqslant t_{\rm ST} \\
\phantom{1}3\,988\,\,{\rm \kms}\,\, \left(\frac{E_{\rm SN}}{E_{51}}\right)^{1/2} \,\,  \left(\frac{M_{\rm ej}}{M_\odot}\right)^{-1/2}  \left(\frac{t}{t_{\rm ST}}\right)^{-3/5} & \text{if } t_{\rm ST} < t \leqslant t_{\rm PD}\\
\phantom{11\,}135\,\,{\rm \kms}\,\, \left(\frac{E_{\rm SN}}{E_{51}}\right)^{0.07} \,\,  \left(\frac{\densini}{n_0}\right)^{0.13} \alpha^{-3/5}\ \left(\frac{t}{t_{\rm PD}}\right)^{-5/7} & \text{if } t_{\rm PD} < t \leqslant t_{\rm MC}\\
\phantom{11\,1}23\,\,{\rm \kms}\,\, \left(\frac{E_{\rm SN}}{E_{51}}\right)^{0.07} \,\,  \left(\frac{\densini}{n_0}\right)^{0.13} \alpha^{-3/5}\ \left(\frac{\beta}{10}\right)^{-5/7} \left(\frac{t}{t_{\rm MC}}\right)^{-3/4}  & \text{if } t_{\rm MC} < t \leqslant t_{\rm fade}
\end{dcases}
\end{equation}
\endgroup
In these expressions, $E_{51} = 10^{51}$ erg and $n_0 = 1$ cm$^{-3}$ are normalization constants. The times $t_{\text{ST}}$, $t_{\text{PD}}$, and $t_{\text{MC}}$ mark the onset of the ST, PD, and MC stages, respectively, $t_{\text{fade}}$ is the time at which the remnant merges with the ambient medium, while $\alpha$ and $\beta$ are scaling factors described below.

\begin{figure}[h!]
\centering
\includegraphics[width = \linewidth ]{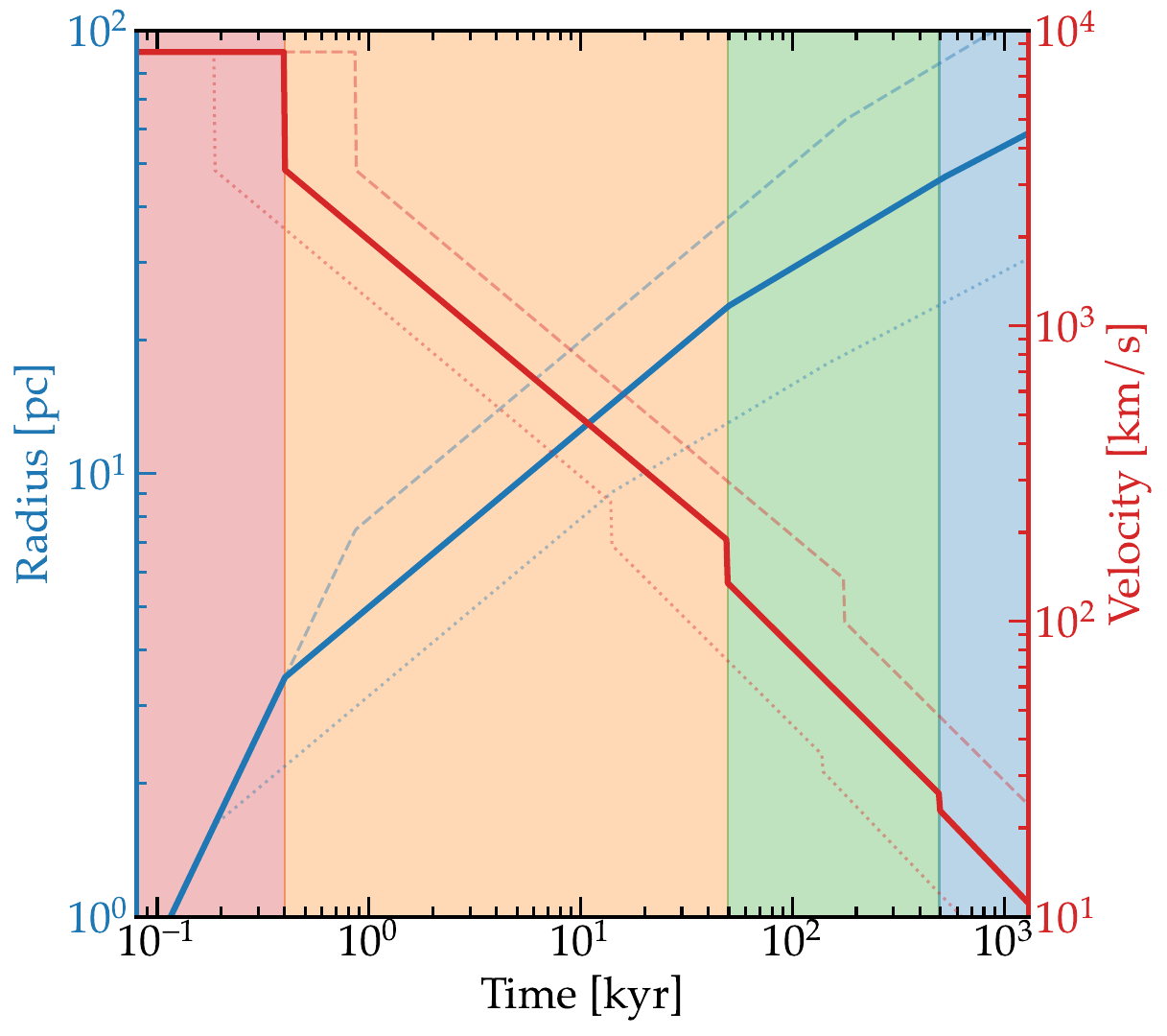}
\caption{Classical piecewise analytical evolutions of the radius and velocity of a one-dimensional spherical SNR \citep[e.g.,][]{Draine2011,Vink2020}. The radius (blue curves) and expansion velocity (red curves) are shown for remnants expanding in a homogeneous medium with proton densities $\densini=0.1$ (dashed), 1 (solid), and 10~\cc\ (dotted). All other parameters are set to their standard values listed in Table~\ref{tab:param_galac}. For the reference case $\densini = 1~\cc$, the successive FE, ST, PD, and MC stages are highlighted by red, orange, green, and blue shaded areas, respectively.}
\label{fig:four_phases}
\end{figure}

The onset time of the Sedov–Taylor stage is set by imposing continuity of the blast radius between the FE and ST phases,
\begin{equation} \label{eq:tst}
t_{\rm ST} = 0.303\,\,{\rm kyr}\,\, \left(\frac{E_{\rm SN}}{E_{51}}\right)^{-1/2} \,\, \left(\frac{M_{\rm ej}}{M_\odot}\right)^{5/6} \,\, \left(\frac{\densini}{n_0}\right)^{-1/3}.
\end{equation}
Following \citet{Draine2011}, the onset time of the pressure-driven stage is defined as the time when the integrated radiative losses of the blast interior become comparable to its thermal energy ($\Delta E_{\rm th} / E_{\rm th} = -1/3$),
\begin{equation} \label{eq:tpd}
t_{\rm PD} = 49.3\,\, {\rm kyr}\,\, \alpha\ \left(\frac{E_{\rm SN}}{E_{51}}\right)^{0.22} \,\, \left(\frac{\densini}{n_0}\right)^{-0.55}.
\end{equation}
Because this definition of the cooling time is somewhat arbitrary, an additional parameter $\alpha$ is introduced relative to \citet{Godard2024b} in order to allow for shorter or longer ST phases and to facilitate comparison with alternative prescriptions used in the literature. The onset time of the momentum-conserving stage is more uncertain \citep{Kim2015}. For simplicity, it is parameterized here as a multiple of $t_{\rm PD}$, 
\begin{equation} \label{eq:tmc}
t_{\rm MC} = \beta\ t_{\rm PD},
\end{equation}
where $\beta=10$ is adopted as a fiducial value \citep{Godard2024b}. The fading time is finally defined as the epoch at which the terminal shock velocity drops below the 1D velocity dispersion of the ambient medium, $V_{\text{fade}}$,  
\begingroup\makeatletter\def\f@size{9.4}\check@mathfonts
\begin{equation} \label{eq:tfade}
t_{\rm fade} = 1.50\,\, {\rm Myr}\,\, \left(\frac{E_{\rm SN}}{E_{51}}\right)^{0.31} \,\, \left(\frac{\densini}{n_0}\right)^{-0.38} \alpha^{1/5}\ \left(\frac{\beta}{10}\right)^{1/21} \left(\frac{V_{\rm fade}}{V_{30}}\right)^{-4/3},
\end{equation}
\endgroup
where $V_{30} = 30$~\kms\ corresponds to three times the 1D velocity dispersion of the WNM \citep[e.g.,][]{Haud2007}.

Figure~\ref{fig:four_phases} illustrates the piecewise analytical evolution of the blast radius and velocity for SNRs expanding into homogeneous media of different ambient densities. Depending on the surrounding density, SNRs persist for total lifetimes ranging from a few $10^5$ to several $10^6$~yr. While the FE and ST stages govern the early evolution, they represent only a minor fraction of the remnant lifetime. In contrast, the radiative stage (including the PD and MC phases), which is the focus of the present paper, accounts for most of the remnant lifetime, with duration at least an order of magnitude longer than the preceding stages. The expansion velocity of radiative SNRs ranges from 30 to a few hundreds \kms. As the expansion slows monotonically with time, either the remnant age or the terminal shock velocity can be used interchangeably as a measure of its evolutionary stage.

\subsection{Energy budget and partition}

The analytical evolution of the terminal shock radius and velocity can be used to estimate the energy budget of the SNR and its partition between kinetic and thermal components that are eventually dissipated during the radiative stages. The total kinetic energy dissipated by the terminal shock during the PD and MC phases is
\begin{equation}
E_{\text{K}} = \int_{t_{\text{PD}}}^{\infty} \frac{1}{2} \rho V_{\text{B}}^3 (t)\ 4 \pi R_{\text{B}}^2 (t)\ dt,
\end{equation}
where 
\begin{equation}
\rho = 1.4 m_{\rm H} \densini
\end{equation}
is the mass density of the ambient medium. Using Eqs.~\ref{eq:rb} and \ref{eq:vb} yields 
\begin{equation} \label{eq:ekn}
E_{\text{K}} = 10^{50}  ~\text{erg} ~\left[5.2\  (1 - \beta^{-4/7} ) + 2.8\ \beta^{-4/7}\right] \left(\frac{E_{\rm SN}}{E_{51}}\right).
\end{equation}
Because the FE and ST stages are adiabatic, the remaining energy is stored as thermal energy in the hot interior at the onset of the radiative phase. The total thermal energy radiated during the PD and MC stages, deduced from energy conservation, is therefore
\begin{equation} \label{eq:eth}
E_{\text{th}} = 10^{50}  ~\text{erg}\ \left[4.8 + 2.4\ \beta^{-4/7} \right]  \left(\frac{E_{\rm SN}}{E_{51}}\right).
\end{equation}

Taken together, Eqs.~\ref{eq:ekn} and \ref{eq:eth} show that the energy partition is controlled solely by the parameter $\beta$. If $\beta=1$, which corresponds to the absence of a PD stage, about $\sim 72$\% of the supernova energy is in the form of thermal energy in the hot interior at the end of the ST phase, while the remaining $\sim 28$\% is carried by the kinetic energy of the expanding shell. If $\beta>1$, the pressure of the hot interior supplies momentum and a fraction of its thermal energy is converted into kinetic energy of the shell. In the limit $\beta \rightarrow \infty$, $E_{\rm K} \rightarrow 52$\% and $E_{\rm th} \rightarrow 48$\%, while for the fiducial value $\beta=10$, $E_{\rm K} \sim 46$\% and $E_{\rm th} \sim 54$\%. Simple energy budget arguments therefore show that both the hot interior and the forward radiative shock carry a significant fraction of the supernova energy, regardless of the value of $\beta$, and must be taken into account to model the radiative evolution of SNRs, their impact on the surrounding medium, and their observational tracers.

\subsection{Initial conditions of the hot bubble}
\label{app:initial_hb}

As discussed in the main text (Sect.~\ref{sec:hotbb}), the hot interior of the remnant is modeled as a homogeneous spherical shell, referred to as the hot bubble, in pressure equilibrium with the ram, thermal, and magnetic pressures exerted by the surrounding medium. Within this simplified framework, the thermodynamical properties of the hot bubble at the onset of the PD stage can be derived from conservation arguments.

The total number of particles contained in the hot bubble at the onset of the pressure-driven stage is taken to be equal to the number of particles swept up during the FE and ST phases, yielding
\begin{equation}
N =  3.8 \times 10^{60}\ \alpha^{6/5}\ \left(\frac{E_{\rm SN}}{E_{51}}\right)^{0.87}  \left(\frac{\densini}{n_0}\right)^{-0.26}.
\end{equation}
Given the thermal energy at the onset of the radiative phase (Eq.~\ref{eq:eth}), the corresponding initial temperature of the hot bubble is
\begin{equation} \label{eq:thb}
T = 10^5\ \text{K}\ \alpha^{-6/5}\ \left[6.1 + 3.1\ \beta^{-4/7} \right]\ \left(\frac{E_{\rm SN}}{E_{51}}\right)^{0.13} \left(\frac{\densini}{n_0}\right)^{0.26}.
\end{equation}
It follows that the thermal pressure of the hot bubble can be written as
\begin{equation} \label{eq:pth}
P_{\text{th}} = 10^{-10}\  {\rm erg}\ {\rm cm}^{-3}\ \eta\ \alpha^{-6/5}\ \left[2.0 + 1.0\ \beta^{-4/7}\right] ~ \left(\frac{E_{\rm SN}}{E_{51}}\right)^{0.13} \left(\frac{\densini}{n_0}\right)^{1.26},
\end{equation}
where $\eta$ is the density contrast between the homogeneous hot bubble and the surrounding ambient medium. Using Eqs.~\ref{eq:rb} and \ref{eq:vb}, the ram pressure at the onset of the PD phase is
\begin{equation} \label{eq:pram}
P_{\text{ram}} =4.3 \times 10^{-10}\ {\rm erg}\ {\rm cm}^{-3}\ \alpha^{-6/5}\ \left(\frac{E_{\rm SN}}{E_{51}}\right)^{0.14} \left(\frac{\densini}{n_0}\right)^{1.26}.\\
\end{equation}
Imposing pressure equilibrium finally leads to
\begin{equation} \label{eq:etahb}
\eta = \frac{4.3}{2+\beta^{-4/7}}
\end{equation}
and an initial proton density
\begin{equation} \label{eq:nhb}
n_{\rm H} = \eta\ \densini.
\end{equation}
Equations~\ref{eq:thb}, \ref{eq:etahb}, and \ref{eq:nhb} define the set of initial conditions of the hot bubble. Its subsequent evolution is computed assuming pressure balance,
\begin{equation}
P(t) = P_{\rm ram}(t) + P_{\rm th}^0 + P_{\rm mag}^0,
\end{equation}
where $P(t)$ is the total (thermal plus magnetic) pressure of the hot bubble, $P_{\rm ram}(t)$ is the time-dependent ram-pressure, and $P_{\rm th}^0$ and $P_{\rm mag}^0$ are the thermal and magnetic pressures of the ambient medium.

\section{Dilution of the radiation field}
\label{app:dilution}

As described in Sect.~\ref{sec:SNR_model} (see Fig.~\ref{fig:onion}), the ionizing photons that irradiate the pre-shock gas are emitted by the post-shock layer and the hot bubble, i.e. a thin shell located behind the shock front. Because this shell is not a point source and has a finite angular size, the ionizing flux seen from a given point of the surrounding medium is not given by a simple inverse-square law. In this appendix, we derive the limb brightening and geometric dilution factor applied to the shell emission.

\subsection{Geometry}
\begin{figure}[h!]
\centering
\includegraphics[width = \linewidth*5/6 ]{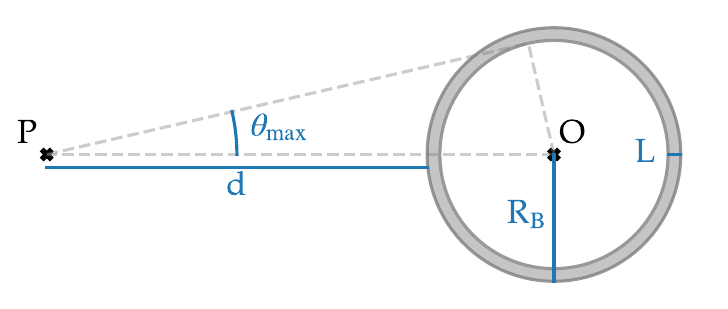}
\caption{ Geometry used to compute the dilution of the radiation field emitted by a r-SNR of radius $R_B$. The gray annulus corresponds to the emitting shell of thickness $L$. The black cross marks a point $P$ of the pre-shock medium, at a distance $d$ from the shock front. The angle $\theta_{\rm max}$ is the angular radius under which the inner edge of the shell is seen from $P$. We consider that only the lines of sight within $\theta_{\rm max}$ contribute to the ionizing flux (see Eq.~\ref{eq:Fint}).}
\label{fig:dilution}
\end{figure}

The emitting region is modeled as a spherical shell of radius $R_B$ and thickness $L$ (see Fig.~\ref{fig:dilution}). We consider a point $P$ of the pre-shock medium located at a distance $d$ from the shock front, i.e. at a distance $R_B+d$ from the center of the remnant. Let $\theta$ be the angle between a given line of sight and the line joining $P$ and the center of the remnant, and define $\mu=\cos\theta$.

Under the approximation that $L \ll R_B$, the narrow annulus of grazing rays that clip the shell without reaching the cavity has a negligible contribution to the ionizing flux. Seen from $P$, the shell therefore covers the directions $\theta\le\theta_{\rm max}$, where $\theta_{\rm max}$ is the angular radius of the inner edge and verifies
\begin{equation}
    \mu_{\rm min}=\cos\theta_{\rm max} =\frac{\sqrt{(R_B+d)^{2}-(R_B-L)^{2}}}{R_B+d}.
    \label{eq:mumin}
\end{equation}

\subsection{Ionizing flux}
\label{app:dilution_flux}

We define $\tau$ as the optical depth of the preshock medium between $P$ and the shock front. Because $L \ll R_B$, a line of sight of inclination $\theta$ crosses an emitting column $1/\mu$ longer than for $\theta=0$. The intensity along that direction is therefore approximated by
\begin{equation}
    I(\mu)=\frac{I_{0}}{\mu}\,e^{-\tau/\mu},
    \label{eq:Imu}
\end{equation}
where $I_{0}$ is the radial ($\theta=0$) surface brightness of the shell, the $1/\mu$ factor is the limb-brightening term, and $e^{-\tau/\mu}$ accounts for the absorption by the surrounding gas.

The ionizing flux at $P$ therefore writes
\begin{equation}
    F=2\pi\int_{\mu_{\rm min}}^{1}\frac{I_{0}}{\mu}\,e^{-\tau/\mu}\mathrm{d}\mu .
    \label{eq:Fint}
\end{equation}
Substituting $u=\tau/\mu$, Eq.~\ref{eq:Fint} integrates to
\begin{equation}
    F=2\pi I_{0}\int_{\tau}^{\tau/\mu_{\rm min}}\frac{e^{-u}}{u}~ \mathrm{d}u
     =2\pi I_{0}\left[E_{1}(\tau)-E_{1}\left(\tau/\mu_{\rm min}\right)\right],
    \label{eq:Ffinal}
\end{equation}
where $E_{1}(x)=\int_{x}^{\infty}e^{-t}/t ~\mathrm{d}t$ is the first exponential integral.

\end{document}